\documentclass[12pt,nofootinbib,superscriptaddress]{revtex4}

\usepackage{amssymb,color,paralist,amsmath,epsfig}
\usepackage{graphicx}
\usepackage{comment}
\usepackage{amsfonts}
\usepackage{color}

\usepackage{relsize}

\usepackage[colorlinks,citecolor=blue,linktoc=all,linkcolor=cyan]{hyperref}

\newcommand{\beq}{\begin{equation}}
\newcommand{\eeq}{\end{equation}}

\newcommand{\ber}{\begin{eqnarray}} 
\newcommand{\eer}{\end{eqnarray}}

\begin{document}

\title{Two-photon exchange correction in elastic unpolarized electron-proton scattering at small momentum transfer}

\author{O. Tomalak}
\email[e-mail: ]{tomalak@uni-mainz.de}
\affiliation{Institut f\"ur Kernphysik, Johannes Gutenberg Universit\"at, D-55099 Mainz, Germany}
\affiliation{PRISMA Cluster of Excellence, Johannes Gutenberg-Universit\"at, D-55099 Mainz, Germany}
\affiliation{Department of Physics, Taras Shevchenko National University of Kyiv, UA-01601 Kyiv Ukraine}
\author{M. Vanderhaeghen}
\affiliation{Institut f\"ur Kernphysik, Johannes Gutenberg Universit\"at, D-55099 Mainz, Germany}
\affiliation{PRISMA Cluster of Excellence, Johannes Gutenberg-Universit\"at, D-55099 Mainz, Germany}

\date{\today}

\begin{abstract}
We evaluate the two-photon exchange (TPE) correction to the unpolarized elastic electron-proton scattering at small momentum transfer $ Q^2 $. We account for the inelastic intermediate states approximating the double virtual Compton scattering by the  unpolarized forward virtual Compton scattering. The unpolarized proton structure functions are used as input for the numerical evaluation of the inelastic contribution. Our calculation reproduces the leading terms in the $ Q^2 $ expansion of the TPE correction and goes beyond this approximation by keeping the full $ Q^2 $ dependence of the proton structure functions. In the  range of small momentum transfer our result is in good agreement with the empirical TPE fit to existing data.
\end{abstract}

\maketitle

\tableofcontents

\section{Introduction}
\label{sec1}

The discrepancy in the ratio of the electric ($ G_{E} $) over magnetic ($ G_{M} $) proton form factor (FF) extractions, when comparing the unpolarized scattering using the Rosenbluth separation technique \cite{Rosenbluth:1950yq} with the scattering of a polarized electron beam followed by the recoiling proton's polarization measurement \cite{Jones:1999rz, Gayou:2001qd,Punjabi:2005wq,Puckett:2010ac}, or the scattering on a polarized proton target \cite{Jones:2006kf}, spurred a reconsideration of the formalism of elastic electron-proton scattering. The approximation of the exchange of one photon (OPE) fails at larger momentum transfers, when the relative contribution of the processes with exchange of two photons (TPE) increases \cite{Guichon:2003qm,Blunden:2003sp}; see e. g. Refs. \cite{Carlson:2007sp,Arrington:2011dn} for reviews. During the last decade, TPE corrections were estimated in different approaches \cite{Blunden:2003sp,Chen:2004tw,Borisyuk:2008es,Borisyuk:2008db,Kivel:2009eg,Kivel:2012vs}. Few direct measurements of the TPE corrections were proposed and performed at Jlab \cite{Meziane:2010xc,Adikaram:2014ykv}, at VEPP-3 \cite{Rachek:2014fam}, and at DESY (OLYMPUS experiment) \cite{Milner:2013daa}, with the analysis of the latter currently in progress.

Besides the importance of TPE corrections at larger momentum transfers, when extracting the $ G_E / G_M $ ratio, recent high precision measurements of elastic electron-proton scattering at lower momentum transfer also make it necessary to better quantify the TPE effects in order to extract proton FFs. Despite the relatively small absolute correction to the electric and Zemach radii \cite{Blunden:2005jv}, TPE corrections can significantly change the magnetic radii \cite{Bernauer:2013tpr}, and therefore this correction has to be additionally studied. In this context, several improved evaluations of TPE box contributions with different intermediate states were performed recently for elastic electron-proton scattering \cite{Borisyuk:2012he,Borisyuk:2013hja,Lorenz:2014yda,Borisyuk:2015xma,Tomalak:2014sva,Zhou:2014xka}.

The importance of the qualitative and quantitative understanding of TPE effects is also necessary in view of the 4-7 standard deviations discrepancy in the proton rms charge radius extraction when comparing muonic hydrogen spectroscopy extractions \cite{Pohl:2010zza, Antognini:1900ns} with experiments using electrons \cite{Mohr:2012tt,Bernauer:2013tpr,Lorenz:2014yda,Lee:2015jqa,Arrington:2015ria}; see Ref. \cite{Carlson:2015jba} for a recent review. The TPE corrections constitute the leading hadronic uncertainties in the extraction of the proton charge radius from muonic hydrogen spectroscopy. Several recent works \cite{Nevado:2007dd,Hill:2011wy,Carlson:2011zd,Birse:2012eb,Alarcon:2013cba,Peset:2014jxa} estimated that they account for $ 10\% $ - $ 15\% $ of the correction needed to reconcile the proton charge radius extractions from muonic hydrogen spectroscopy and electron experiments. 

In the region of very small momentum transfer $ Q^2 $, the TPE corrections can be determined model independently. Consequently, each TPE calculation should give the same result or at least reproduce some characteristic features in the low-$Q^2$ limit. The leading term in the momentum transfer expansion  ($ Q^2 $) arises from the scattering cross section of the relativistic massless electron on a point charged target in Dirac theory given by the so-called Feshbach correction \cite{McKinley:1948zz}. This result was reproduced later by Brown in Ref. \cite{Brown:1970te} as the leading term in the expansion of the TPE correction with the proton intermediate state (elastic TPE). Brown also found that the subleading $ Q^2 \ln^2 Q^2 $ term entirely arises from the elastic TPE, while the subleading $ Q^2 \ln Q^2 $ term arises from both the elastic and inelastic TPEs. In Ref. \cite{Brown:1970te} it was shown that the inelastic TPE correction to the $ Q^2 \ln Q^2 $ term can be expressed in terms of an energy integral over the total photoabsorption cross section on a proton target. A recent numerical estimate of this term was given in Ref. \cite{Gorchtein:2014hla}.

In this work, we extend the low-$Q^2$ limit of the inelastic TPE contribution beyond the leading $ Q^2 \ln Q^2 $ term. We express its contribution in terms of the two unpolarized forward virtual Compton scattering amplitudes. In detail, the TPE correction is presented as a double integral over the virtual photon energy and the virtuality of the unpolarized nucleon structure functions. Our results reproduce the known $ Q^2 \ln Q^2 $ term of Ref. \cite{Brown:1970te} and get rid of the hadronic scale $ \Lambda $ dependence, which arises to the order $ Q^2 \ln Q^2 $, thus partially accounting for the $ Q^2 $ term and higher terms in the momentum transfer expansion. We keep the $ Q^2 $ dependence in the proton structure functions in order to reproduce the model-independent TPE correction at low momentum transfers, as well as to provide numerical estimates in the most accurate way. 

The paper is organized as follows. We introduce the general description of the TPE correction to the unpolarized elastic electron-proton scattering in terms of the double virtual Compton scattering process in Sec. \ref{sec2} and reproduce the leading terms in the momentum transfer expansion of the elastic TPE contribution in Sec. \ref{sec3}.  We express the leading term of the inelastic TPE correction as an integral over the unpolarized proton structure functions in Sec. \ref{sec4}. We show in Sec. \ref{sec5} that the subtraction function in the forward Compton scattering amplitude $ T_1 $ is negligible for elastic electron-proton scattering. In Sec. \ref{sec6} we describe the details of our evaluation of the inelastic TPE. In Sec. \ref{sec7} we show that our expression reproduces the known result for the leading $ Q^2 \ln Q^2 $ inelastic TPE contribution. We present the results of the numerical evaluation beyond the leading terms in Sec. \ref{sec8}. Our conclusions and an outlook are given in Sec. \ref{sec9}.

\section{TPE correction in terms of VVCS}
\label{sec2}

\begin{figure}[h]
\begin{center}
\includegraphics[width=.34\textwidth]{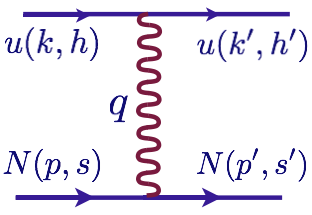}
\end{center}
\caption{Elastic electron-proton scattering in the OPE approximation.}
\label{OPE_kinematics}
\end{figure}

The elastic scattering of an electron off a proton target in the OPE approximation is described by the helicity amplitude,
\ber
 T^{1 \gamma}  = \frac{e^2}{Q^2} \bar{u}\left(k',h'\right) \gamma_\mu u \left(k,h\right) \cdot \bar{N} \left(p',s'\right) \Gamma^\mu \left( Q^2 \right) N \left(p,s \right),
\eer
with the electron (proton) initial and final momenta given by $ k $, $ k'$ ($ p $, $ p' $) respectively, the squared momentum transfer $ Q^2 = - q^2 =  - \left( k - k^\prime \right)^2 $, and the unit of electric charge $ e $. The kinematics of the process and the electron (proton) initial and final helicities $ h,h^\prime $ ($ s, s^\prime $) are shown in Fig. \ref{OPE_kinematics}.

The electromagnetic vertex for the transition $\gamma^* \left(q\right) + N\left(p\right) \to N\left(p + q\right)$ is given by the Lorentz and gauge invariant form that preserves parity and charge conjugation symmetries:
\beq
\label{emvertex}
\Gamma^\mu \left( Q^2 \right) \;=\; F_D\left(Q^2\right) \, \gamma^\mu \;+\;
F_P\left(Q^2\right) \, \frac{i \sigma^{\mu \nu} q_\nu}{2 M} \, ,
\eeq
with M the proton mass and $F_D$ and $F_P$ the Dirac and Pauli FFs of the proton,
normalized to $F_D \left(0\right)=1$ and $F_P\left(0\right)=\kappa$,
where $\kappa$ is 
the anomalous magnetic moment in units of $e/ \left(2M\right)$.  
The Dirac and Pauli FFs are often equivalently expressed in terms of the Sachs magnetic and electric FFs as
\ber  \label{Sachs_ffs}
 G_M  =  F_D + F_P , ~~~~~~~
 G_E  = F_D - \tau F_P,
\eer 
with $ \tau = Q^2 / \left(4 M^2\right)  $.

\begin{figure}[h]
\begin{center}
\includegraphics[width=.35\textwidth]{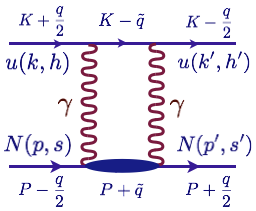}
\end{center}
\caption{Direct and crossed TPE diagrams.}
\label{TPE_kinematics}
\end{figure}

The exchange of two photons contributes to the elastic electron-proton scattering through the TPE diagram shown in Fig. \ref{TPE_kinematics}.

\begin{figure}[h]
\begin{center}
\includegraphics[width=.35\textwidth]{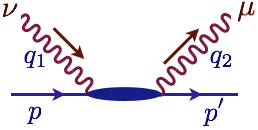}
\end{center}
\caption{Nonforward VVCS process.}
\label{VVCS}
\end{figure}

The lower blob in Fig. \ref{TPE_kinematics} is given by the double virtual Compton scattering (VVCS) process on a proton (see Fig. \ref{VVCS}): $ \gamma^\ast \left(q_1, \lambda_1\right) + N\left(p,s\right) \to \gamma^\ast\left(q_2, \lambda_2\right) + N\left(p^\prime, s^\prime\right) $. The VVCS amplitude $ T_{\lambda_2 s', \lambda_1 s} $ can be written in terms of the VVCS tensor $ M^{\mu \nu} $ as
\ber
T_{\lambda_2 s', \lambda_1 s} = \varepsilon_\nu\left(q_1, \lambda_1\right) \varepsilon^\ast_\mu\left(q_2, \lambda_2\right) \cdot  \bar{N}\left(p',s'\right) M^{\mu \nu} N\left(p,s\right),
\eer
where $ \varepsilon_\nu, ~\varepsilon^\ast_\mu $ denote the virtual photon polarization vectors, $ N, \bar{N} $ the proton spinors, and  $ \lambda_1, \lambda_2 ~(s, s^\prime) $ the photon (proton) helicities. The initial and final virtual photon momenta $ q_1$ and $q_2$ are related to the momentum transfer $ q $ and the loop variable $ \tilde{q} $ in Fig. \ref{TPE_kinematics} by
\ber
q_1 = \tilde{q} + \frac{q}{2}, \qquad q_2 = \tilde{q} - \frac{q}{2}.
\eer

For the following, it is also convenient to introduce the averaged lepton (nucleon) four-momenta $ K $ ($ P $), respectively, as
\ber
K = \frac{1}{2} \left( k + k' \right), \quad P = \frac{1}{2} \left( p + p' \right).
\eer

The TPE correction $ \delta_{2 \gamma } $ to the unpolarized $ e^- p $ scattering cross section, due to the graph in Fig. \ref{TPE_kinematics}, in the leading order of the fine structure constant $ \alpha = e^2 /\left(4 \pi\right) $ can be defined as the difference between the measured cross section $ \sigma $ and the cross section in the OPE approximation $ \sigma_{1 \gamma} $ as
\ber
\sigma \simeq \sigma_{1 \gamma} \left( 1 + \delta_{2 \gamma} \right).
\eer
The TPE correction $ \delta_{2 \gamma}$ is given by the interference of the one-photon exchange $ T^{1 \gamma} $ and the two-photon exchange $ T^{2 \gamma}$ amplitudes,
\ber \label{TPE_correction}
\delta_{2 \gamma} = \frac{2 \Re \left(\sum \limits_{spin}T^{2 \gamma} \left( T^{1 \gamma} \right)^* \right)}{\sum \limits_{spin}  |T^{1 \gamma}|^2},
\eer
with the interference term
\ber
& & \hspace{1.2cm} 2 \Re \left(\sum \limits_{spin}T^{2 \gamma} \left( T^{1 \gamma} \right)^* \right)  =  -  \Re  \frac{4 \pi  e^4}{Q^2}  \mathop{\mathlarger{\int}} \frac{ i \mathrm{d}^4 \tilde{q}}{\left( 2 \pi \right)^4} \frac{L^{\mu \nu \alpha} H_{\mu \nu \alpha}  }{\left(\left(\tilde{q} - \frac{q}{2} \right)^2 - \mu^2\right)\left(\left(\tilde{q} + \frac{q}{2} \right)^2 - \mu^2\right)} , \nonumber \\
& & \hspace{0.2cm} L^{\mu \nu \alpha}  =  \textbf{Tr}  \left\{  \left( \gamma^\mu \frac{\hat{K}-\hat{\tilde{q}}+m}{\left(K-\tilde{q}\right)^2-m^2} \gamma^\nu + \gamma^\nu \frac{\hat{K}+\hat{\tilde{q}}+m}{\left(K+\tilde{q}\right)^2-m^2} \gamma^\mu  \right) \left(\hat{k}+m\right) \gamma^\alpha \left(\hat{k^\prime}+m\right) \right\}, \nonumber \\
& & \hspace{4.2cm} H^{\mu \nu \alpha}  =  \textbf{Tr} \left\{ M^{\mu \nu} \left(\hat{p}+M \right) \Gamma^\alpha \left( Q^2 \right) \left(\hat{p^\prime}+M \right)   \right\}, \nonumber 
\eer
where $ m $ denotes the lepton mass, and where $ \mu $ denotes the small photon mass, that plays a role of the IR regulator.

Furthemore, the VVCS tensor $ M^{\mu \nu} $ reads, in the notation of Refs.~\cite{Drechsel:1997xv,Drechsel:1998zm}\footnote{Note, however, that our convention for the photon field four-momenta indices $ \mu $ and $ \nu $ (see Fig. \ref{VVCS}) is opposite to Refs. \cite{Drechsel:1997xv}, \cite{Drechsel:1998zm}, which changes the sign of the tensors that are antisymmetric under $ \mu \leftrightarrow \nu $. }, as
\ber 
M^{\mu \nu} &=& \alpha \sum_{i \in J} B_i \left(q_1^2, q_2^{2}, q_1 \cdot q_2, \tilde{q} \cdot P\right) \, T_i^{\mu \nu} , 
\nonumber  \\
&&J = \left \{ 1,..., 21 \right \} \backslash \left\{5, 15, 16\right \},
\label{eq:dvcs}
\eer
where the 18 independent tensors $T_i^{\mu \nu}$ were constructed to be gauge invariant, 
and free of kinematical singularities and constraints, following the procedure outlined in Ref. \cite{Tarrach:1975tu}. The invariant amplitudes $ B_i $ satisfy definite transformation properties with respect to photon crossing as well as charge conjugation combined with proton crossing as detailed in Refs.~\cite{Drechsel:1997xv,Drechsel:1998zm}. The lack of knowledge of the amplitudes $ B_i $ does not allow one to evaluate the TPE correction in the general case.

In this work, we will consider TPE correction to the unpolarized elastic electron-proton scattering in the low-$Q^2$ limit. The momentum transfer accessed by current electron-proton scattering experiments is much larger than the electron mass $ m $.  We will consider the kinematic range $ m^2 \ll Q^2 \ll M^2,~ M E $.

\section{Elastic TPE contribution at small $ Q^2 $}
\label{sec3}

We start our study by checking the low-$Q^2$ limit for the case of proton intermediate states in the lower blob of the TPE graph; see Fig. \ref{TPE_kinematics}. In this case, we can compare our result with the known result for the elastic contribution. 

The leading terms of the low-$ Q^2 $ expansion of the elastic contribution to the TPE correction $ \delta^{Born}_{2 \gamma} $ in the unpolarized elastic $ e^{-} p $ scattering are given by the sum of the IR divergent piece $ \delta^{IR}_{2 \gamma} $, the Feshbach term $ \delta_F $ \cite{McKinley:1948zz} expanded to the leading order, and the finite logarithmic corrections as
\ber \label{ep_elastic_forward}
\delta^{Born}_{2 \gamma} & \to & \delta^{IR}_{2 \gamma} + \delta_F  +  \frac{\alpha}{\pi} 
 \frac{Q^2}{M E} 
\ln \left( \frac{Q}{2 E} \right) \left[ \ln \left( \frac{Q}{2 E} \right) + 1 \right] + \mathrm{O} \left(\frac{Q^2}{M^2}, \frac{Q^2}{M E}, \frac{Q^2}{E^2}  \right),
\eer
where $ \delta_F $ and $ \delta^{IR}_{2 \gamma} $ are given by
\ber
\delta_F \to \frac{ \alpha \pi Q}{2 E}, \quad \quad \quad \delta^{IR}_{2 \gamma} \to \frac{\alpha }{\pi} \frac{Q^2}{M E} \ln \frac{\mu^2}{Q^2}.
\label{deltaIR_Feshbach_corr}
\eer

This result was obtained in two equivalent ways in the literature: either by considering the low-$Q^2$ limit of the box graph or by using dispersion relations for the invariant amplitudes of the elastic electron-proton scattering; see Refs. \cite{Brown:1970te,Tomalak:2014sva} for details.

The correction of Eq. (\ref{ep_elastic_forward}) is given by the graph with two Dirac structures $ \gamma^\mu $ in the photon-proton-proton vertices, which are independent of the proton structure. The VVCS tensor $ M^{\mu \nu}_{point} $ in this case is given by the sum of three nonvanishing terms \cite{Tarrach:1975tu,Drechsel:1997xv,Drechsel:1998zm}:
\ber
M^{\mu \nu}_{point} = \alpha \left(B_2^{point} T^{\mu \nu}_{2} + B_{10}^{point} T^{\mu \nu}_{10} + B_{17}^{point} T^{\mu \nu}_{17} \right).
\eer
The amplitudes $ B_2^{point}, ~B_{10}^{point},~ B_{17}^{point} $ and the expressions for the corresponding tensors are given in Appendix \ref{elastic_F1F1_structures}.
 
The leading elastic contribution to the TPE correction of Eq. (\ref{TPE_correction}) is given by the contribution from the unpolarized VVCS amplitude $ B_2^{point} $. Neglecting the subleading terms in the momentum transfer expansion (i.e. taking the limit $ Q^2 \ll M^2,~ M E,~E^2$), the leading terms in the $ Q^2 $ expansion for a Dirac point particle can then be obtained as
\ber \label{lp_elastic_integral}
\delta_{2 \gamma}^{point} \to \frac{16 Q^2 e^2}{M E}  && \mathop{\mathlarger{\int}}  \frac{  i \mathrm{d}^4 \tilde{q}}{\left( 2 \pi \right)^4} \Pi_{P}^{+} \Pi_{P}^{-}\Pi_{K}^{+} \Pi_{K}^{-}\Pi_{Q}^{+} \Pi_{Q}^{-}  \left \{  \left(2 \tilde{q}^2 + \frac{3}{2} Q^2\right) \left(K\cdot P\right) \left(K\cdot \tilde{q}\right) \left(P\cdot\tilde{q}\right) \right.  \nonumber \\
&&  \left. - 2 \left(K\cdot\tilde{q}\right)^2 \left(P\cdot\tilde{q}\right)^2 - \left(\tilde{q}^2 + \frac{Q^2}{4}\right) \left(M^2\left(K\cdot\tilde{q}\right)^2 +  \frac{Q^2}{2} \left(P\cdot\tilde{q}\right)^2\right)    \right.  \nonumber \\
&&   \left. - \left(\tilde{q}^2 + \frac{Q^2}{4}\right)^2 \left(K \cdot P \right)^2 + \frac{Q^2}{2 M E}  \left(K\cdot\tilde{q}\right) \left(P\cdot\tilde{q}\right)^3  \right \},
\eer
where we have introduced the propagator notations:
\ber
 \Pi_{P}^{\pm} = \frac{1}{\left(P\pm \tilde{q}\right)^{2} - M^2}, ~~~~~\Pi_{K}^{\pm} = \frac{1}{\left(K\pm \tilde{q}\right)^{2} - m^2},~~~~~\Pi_{Q}^{\pm} = \frac{1}{\left(\tilde{q} \pm q/2\right)^{2} - \mu^2}.
\eer
The expression in Eq. (\ref{lp_elastic_integral}) reproduces all terms in the low-$Q^2$ expansion of Eq. (\ref{ep_elastic_forward}). The TPE correction from the spin-dependent structures $ B_{10}^{point} T^{\mu \nu}_{10} $ and $B_{17}^{point} T^{\mu \nu}_{17}$ expansion starts with higher powers in $ Q^2 $ in comparison with the expansion of the unpolarized structure $ B_2^{point} T^{\mu \nu}_{2} $ contribution to TPE, and consequently the leading $ Q^2 $ term in the expansion coming from the graph with two vertices $ \gamma_\mu $ is reproduced in a correct way with Eq. (\ref{lp_elastic_integral}). As an example, we describe the derivation of the Feshbach correction [the second term in Eq. (\ref{ep_elastic_forward})] in Appendix \ref{feshbach_and_IR_term}. 

\section{Inelastic TPE correction in terms of the forward virtual Compton scattering amplitudes}
\label{sec4}

The VVCS amplitudes $ B_1,~B_2,~B_3,~B_4,~B_{19} $ contribute to the unpolarized forward virtual Compton scattering tensor. 

The elastic intermediate state contribution from the unpolarized VVCS amplitudes is described by two nonvanishing amplitudes,
\ber
B_1^{Born} & = & - \frac{1}{M}  \frac{\tilde{\nu}^2}{ \left( q_1 \cdot q_2 \right)^2 - \left(2 M \tilde{\nu} \right)^2 } \left\{ F_P \left(Q^2_1\right) F_P \left(Q^2_2\right) \right. \nonumber \\
&& \left.+ \frac{ \left(q_1\cdot q_2 \right) }{\tilde{\nu}^2} \left( F_D \left(Q^2_1\right) F_P \left(Q^2_2\right) + F_P \left(Q^2_1\right) F_D \left(Q^2_2\right) + F_P \left(Q^2_1\right) F_P \left(Q^2_2\right) \right) \right \}, \nonumber \\ \\
B_2^{Born} & = & - \frac{1}{M} \frac{1}{ \left( q_1 \cdot q_2 \right)^2 - \left(2 M \tilde{\nu} \right)^2} \left\{ F_D \left(Q^2_1\right) F_D \left(Q^2_2\right) - \frac{ \left(q_1\cdot q_2 \right)}{4 M^2} F_P \left(Q^2_1\right) F_P \left(Q^2_2\right) \right\},
\eer
with $ Q_1^2 \equiv - q^2_1, ~ Q_2^2 \equiv - q^2_2 $, the kinematic relation $ \left(q_1\cdot q_2 \right) = - \left( \tilde{Q}^2 - \frac{Q^2}{4} \right) $, and notations
\ber
 M \tilde{\nu} = \left( P \cdot \tilde{q} \right), \qquad \tilde{Q}^2 = - \tilde{q}^2.
\eer
These amplitudes contribute to the VVCS tensor of Eq. (\ref{eq:dvcs}) with the following tensor structures:
\ber
T^{\mu \nu}_1 & = &   - \left(q_1\cdot q_2\right) g^{\mu\nu}+q_1^{\mu}q_2^{\nu}, \\
T^{\mu \nu}_2 & = & - 4 \left(P\cdot \tilde{q}\right)^2 g^{\mu \nu} - 4 \left(q_1\cdot q_2\right) P^\mu P^\nu + 4 \left(P\cdot\tilde{q}\right) \left(P^{\nu} q^{\mu}_1 + P^{\mu} q^{\nu}_2\right).
\eer
The other unpolarized amplitudes vanish in the Born approximation, i.e.
\ber
B_3^{Born} = 0, \qquad B_4^{Born} = 0, \qquad B_{19}^{Born} = 0.
\eer

When evaluating the TPE correction, one can simplify the calculation by using explicitly gauge invariance, i.e. $ q_2^\mu L_{\mu \nu \alpha} = 0$ and $ q_1^\nu L_{\mu \nu \alpha} = 0 $ for the virtual photon momentas $ q_1, ~ q_2$. Consequently, we can use the identities:
\ber
q_1^\mu L_{\mu \nu \alpha } = q^\mu L_{\mu \nu \alpha }, \qquad
q_2^\nu L_{\mu \nu \alpha} = - q^\nu L_{\mu \nu \alpha }.
\eer
Approximating the arguments of proton form factors as: $ Q^2_1 \approx Q^2_2 \approx \tilde{Q}^2 - Q^2/4 $, the  $ B_1^{Born} $ and $ B_2^{Born} $ contributions can be obtained from the effective {\it near-forward} VVCS tensor,
\ber
\label{vvcs_agreement_with_elastic}
M_{\mathrm{Born}}^{\mu \nu}\left(\tilde{\nu},\tilde{Q}^2\right)& = &
 - \left( -g^{\mu\nu}+\frac{q^{\mu} q^{\nu}}{\tilde{Q}^2 - \frac{Q^2}{4} }\right)
\mathrm{T}^{\mathrm{Born}}_1\left(\tilde{\nu}, \tilde{Q}^2 - \frac{Q^2}{4}\right) \nonumber \\
& & - \frac{1}{M^2} \left(P^{\mu}+\frac{ M \tilde{\nu} }{\tilde{Q}^2 - \frac{Q^2}{4} }\,q^{\mu}\right) \left(P^{\nu}-\frac{ M \tilde{\nu} }{\tilde{Q}^2 - \frac{Q^2}{4} }\, q^{\nu} \right) \mathrm{T}^{\mathrm{Born}}_2 \left(\tilde{\nu}, \tilde{Q}^2 - \frac{Q^2}{4}\right) , \nonumber \\
\eer%
with the Born contributions to forward VVCS amplitudes $ \mathrm{T}^{\mathrm{Born}}_1 $ and $ \mathrm{T}^{\mathrm{Born}}_2 $ given by
\ber
\mathrm{T}_1^{\mathrm{Born}} \left( \nu, Q^2 \right)  & = & \frac{\alpha}{M} \left(  \frac{Q^4 G^2_M\left(Q^2\right)}{ \left( Q^2 \right)^2 - 4 M^2 \nu^2 } -F_D^2 \left(Q^2\right) \right), \label{t1_Born} \\
\mathrm{T}_2^{\mathrm{Born}} \left( \nu, Q^2 \right)  & = &  4 M Q^2  \alpha \frac{ F_D^2\left(Q^2\right) + \frac{Q^2}{4 M^2} F^2_P \left(Q^2\right) }{\left( Q^2 \right)^2 - 4 M^2 \nu^2}. \label{t2_Born}
\eer
Note that in Eq. (\ref{vvcs_agreement_with_elastic}) the second argument of the VVCS amplitudes $ \mathrm{T}^{\mathrm{Born}}_{1, 2} $ is given by replacing in Eqs. (\ref{t1_Born}) and (\ref{t2_Born}) $ Q^2 \to \tilde{Q}^2 - Q^2 /4 $. 

The expression of Eq. (\ref{vvcs_agreement_with_elastic}) gives the correct forward limit, when $ q = 0 $. For a Dirac point particle, we checked that Eq. (\ref{vvcs_agreement_with_elastic}) corresponds with the exact result for the unpolarized VVCS tensor when replacing $ F_D \left( Q^2\right) \to 1 $ and $ F_P \left(Q^2\right) \to 0 $ in Eqs. (\ref{t1_Born}) and (\ref{t2_Born}). At low $ Q^2 $, it leads to Eq. (\ref{lp_elastic_integral}) and gives the leading terms in the low-$ Q^2 $ expansion of the elastic contribution to $ \delta_{2 \gamma}$ of Eq. (\ref{ep_elastic_forward}). We call the approximation of Eq. (\ref{vvcs_agreement_with_elastic}) the {\it near-forward} approximation.

We now turn to the inelastic contributions to the unpolarized forward Compton amplitudes $ \mathrm{T}_1, ~\mathrm{T}_2 $. They are expressed in terms of the VVCS invariant amplitudes $ B_i $ in forward kinematics by
\ber
\mathrm{T}_1\left(\tilde{\nu}, \tilde{Q}^2\right) & = & \alpha \left( \tilde{Q}^2 B_1 - 4 M^2 \tilde{\nu}^2 B_2 + \tilde{Q}^4 B_3 - 4 M \tilde{\nu} \tilde{Q}^2 B_4 \right), \\
\mathrm{T}_2 \left(\tilde{\nu}, \tilde{Q}^2\right) & = & 4 M^2 \tilde{Q}^2  \alpha  \left( - B_2 - \tilde{Q}^2 B_{19} \right).
\eer
In this work, we choose the tensor form of Eq. (\ref{vvcs_agreement_with_elastic}) to describe the inelastic TPE correction and only keep the amplitudes $ B_1 $ and $ B_2 $ in the region of small momentum transfer.

We describe the VVCS invariant amplitudes $B_i \left( q_1^2, ~q_2^{2}, ~q_1 \cdot q_2, ~\tilde{q} \cdot P \right)$ as
\ber
B_i \left( q_1^2, q_2^{2}, ~q_1 \cdot q_2, ~\tilde{q} \cdot P \right) \simeq B_i \left( -  \tilde{Q}^2 + \frac{Q^2}{4}, ~ -  \tilde{Q}^2 + \frac{Q^2}{4}, ~ -  \tilde{Q}^2 + \frac{Q^2}{4}, ~ M \tilde{\nu} \right),
\eer
where we use the approximation $ q_1^2 \approx q_2^2 \approx - \tilde{Q}^2 + Q^2/4$ for the VVCS amplitudes.

 The  {\it near-forward} approximation allows one to obtain the first two terms of the inelastic TPE expansion coming from the proton structure functions $ F_1 $ and $ F_2 $:
\ber \label{inelastic_expansion}
\delta^{inel}_{2 \gamma} \approx  a\left(E\right) Q^2 \ln Q^2 +  b\left(E\right)  Q^2 .
\eer
Nevertheless, we keep the $ Q^2 $ dependence in all kinematic factors, but we do not pretend on the validity of our approximations beyond the expansion of Eq. (\ref{inelastic_expansion}) due to the contribution of double virtual Compton amplitudes besides $ B_1 $ and $ B_2 $. The  {\it near-forward} approximation of Eq. (\ref{vvcs_agreement_with_elastic}) is valid only in the region of small momentum transfer $ Q^2 $. In Sec. \ref{sec8} we will explicitly study the range in $Q^2$ over which such expansion provides a good approximation.

We obtain the TPE correction substituting the {\it near-forward} approximation of the VVCS tensor of Eq. (\ref{vvcs_agreement_with_elastic}) into the general expression for the cross section correction $ \delta_{2 \gamma}$ of Eq. (\ref{TPE_correction}):
\ber
\delta_{2 \gamma}^{inel} & = & 4 \pi \frac{\left(1-\varepsilon\right) G_E}{\varepsilon G_E^2 + \tau G^2_M} \frac{2}{M} \Re \nonumber \\
&& \times \mathop{\mathlarger{\int}} \frac{ i \mathrm{d}^4 \tilde{q}}{\left( 2 \pi \right)^4} \Pi_{Q}^{+} \Pi_{Q}^{-} \times \left \{ 2 \left(K\cdot P\right) \left( \left(K\cdot P\right)^2 - \frac{Q^2}{4} P^2 \right) \frac{\Pi_{K}^{-} + \Pi_{K}^{+}}{M^2} \mathrm{T}_2\left(\tilde{\nu}, \tilde{Q}^2-\frac{Q^2}{4}\right)  \right. \nonumber \\
&& \left. + \frac{Q^2}{\tilde{Q}^2-\frac{Q^2}{4}} \left( \left(P\cdot\tilde{q}\right) \left(K\cdot\tilde{q}\right) P^2 - \left(P\cdot\tilde{q}\right)^2 \left(K\cdot P\right) \right) \frac{\Pi_{K}^{-} + \Pi_{K}^{+}}{M^2} \mathrm{T}_2\left(\tilde{\nu}, \tilde{Q}^2-\frac{Q^2}{4}\right)  \right. \nonumber \\
&& \left. + \left( P^2 + \frac{ \left(P\cdot\tilde{q}\right)^2 Q^2}{\left(\tilde{Q}^2-\frac{Q^2}{4}\right)^2 }\right) \left( \left(K\cdot P\right) \left(K\cdot\tilde{q}\right)  - \frac{Q^2}{4} \left(P\cdot\tilde{q}\right)  \right) \frac{ \Pi_{K}^{-} - \Pi_{K}^{+} }{M^2} \mathrm{T}_2\left(\tilde{\nu}, \tilde{Q}^2-\frac{Q^2}{4}\right) \right. \nonumber \\
&& \left. - 2 \left(P\cdot\tilde{q}\right) \frac{ \tilde{Q}^2-\frac{3 Q^2}{4} }{ \tilde{Q}^2-\frac{Q^2}{4}  } \left( \left(K\cdot P\right)^2  -  \frac{Q^2}{4} P^2 \right) \frac{\Pi_{K}^{-} - \Pi_{K}^{+}}{M^2} \mathrm{T}_2\left(\tilde{\nu}, \tilde{Q}^2-\frac{Q^2}{4}\right) \right. \nonumber \\
&& \left. - 2 \frac{\tilde{Q}^2+\frac{Q^2}{4}}{\tilde{Q}^2-\frac{Q^2}{4}}\left( \left(K\cdot P\right) \left(K\cdot\tilde{q}\right)  - \frac{Q^2}{4} \left(P\cdot\tilde{q}\right)  \right) \left( \Pi_{K}^{-} - \Pi_{K}^{+} \right) \mathrm{T}_1\left(\tilde{\nu}, \tilde{Q}^2-\frac{Q^2}{4}\right) \right \},
\eer
with the photon polarization parameter:
\ber
\varepsilon = \frac{16 \left( K \cdot P \right) ^2 - Q^2 \left( Q^2 + 4M^2 \right)}{16 \left( K \cdot P \right) ^2 + Q^2 \left( Q^2 + 4M^2 \right)}. 
\eer

The imaginary parts of the forward virtual Compton amplitudes $ \mathrm{T}_1, ~\mathrm{T}_2 $, by means of the optical theorem, are directly related to the unpolarized proton structure functions $ F_1 $ and $ F_2 $ by
\ber
\Im \mathrm{T}_1  =  \frac{e^2}{4 M} F_1, \qquad
\Im \mathrm{T}_2  =  \frac{e^2}{4 \tilde{\nu}} F_2.
\eer
To obtain the real parts of the forward Compton amplitudes, we express the dispersion relations (DRs) in terms of the invariant mass of the intermediate state variable $ W^2 $ in order for the boundaries of the integration regions not to depend on $ \tilde{Q}^2 $. This allows us to easily interchange the $ W^2 $ integration with the $ \tilde{q} $ integration. The DRs for forward virtual Compton amplitudes are given by
\ber
\mathrm{T}_1 \left(\tilde{\nu}, \tilde{Q}^2-\frac{Q^2}{4}\right) & & =   \mathrm{T}_1\left(0, \tilde{Q}^2-\frac{Q^2}{4}\right) \nonumber \\
& & + \frac{2 }{\pi }  \mathop{\mathlarger{\int}} \limits^{~~ \infty}_{W^2_{thr}} \frac{ e^2 M \tilde{\nu}^2 F_1\left(W^2,\tilde{Q}^2-\frac{Q^2}{4}\right) \mathrm{d} W^2}{\left( W^2 - P^2 + \tilde{Q}^2 \right)  \left( \left( P + \tilde{q} \right)^2 - W^2 + i \varepsilon \right) \left( \left( P - \tilde{q} \right)^2 - W^2 + i \varepsilon \right)} , \nonumber \\
\mathrm{T}_2 \left(\tilde{\nu}, \tilde{Q}^2-\frac{Q^2}{4}\right) & & =  \frac{1 }{\pi}   \mathop{\mathlarger{\int}} \limits^{~~ \infty}_{W^2_{thr}} \frac{ e^2 M F_2 \left(W^2, \tilde{Q}^2-\frac{Q^2}{4}\right)  \mathrm{d} W^2}{\left( \left( P + \tilde{q} \right)^2 - W^2 + i \varepsilon \right) \left( \left( P - \tilde{q} \right)^2 - W^2 + i \varepsilon \right)},
\eer
with the pion-proton inelastic threshold, $ W^2_{thr} = ( M + m_{\pi})^2  \approx 1.15 ~\mathrm{GeV}^2 $, where $ m_{\pi} $ denotes the pion mass. The unsubtracted  DR for the $ \mathrm{T}_1 $ amplitude is divergent due to the Regge behavior of $ F_1 $ structure function. Consequently, we use a once-subtracted dispersion relation for this amplitude with the conventional subtraction point $ \tilde{\nu} = 0 $.

We find the contribution from the subtraction function $ \mathrm{T}_1 \left(0,\tilde{Q}^2-\frac{Q^2}{4} \right) $ to be negligible in electron-proton scattering experiments; see the following Sec. \ref{sec5} for details. The contributions from the unpolarized proton structure functions $ F_1 $ and $ F_2 $ to the TPE correction, which we denote by $  \delta_{2 \gamma}^{F_1} $ and $ \delta_{2 \gamma}^{F_2} $, respectively, are given by
\ber
 \label{TPE_inelastic1} \delta_{2 \gamma}^{F_1} & = & A ~ \Re \mathop{\mathlarger{\int}} \limits^{~~ \infty}_{W^2_{thr}} \mathrm{d} W^2   \mathop{\mathlarger{\int}} \frac{ i \mathrm{d}^4 \tilde{q}}{\left( 2 \pi \right)^4}  \frac{\Pi_{Q}^{+} \Pi_{Q}^{-} B   \left( \Pi_{K}^{-} - \Pi_{K}^{+} \right) \left(P\cdot\tilde{q}\right)^2 F_1\left(W^2,\tilde{Q}^2-\frac{Q^2}{4}\right) }{\left( W^2 - P^2 + \tilde{Q}^2 \right)  \left( \left( P + \tilde{q} \right)^2 - W^2 + i \varepsilon \right) \left( \left( P - \tilde{q} \right)^2 - W^2 + i \varepsilon \right)}, \nonumber \\ \\
 \label{TPE_inelastic2} \delta_{2 \gamma}^{F_2}& = & A ~ \Re \mathop{\mathlarger{\int}} \limits^{~~ \infty}_{W^2_{thr}} \mathrm{d} W^2  \mathop{\mathlarger{\int}} \frac{ i \mathrm{d}^4 \tilde{q}}{\left( 2 \pi \right)^4} \frac{  \Pi_{Q}^{+} \Pi_{Q}^{-} \left ( C \left( \Pi_{K}^{-} + \Pi_{K}^{+} \right)  + D   \left( \Pi_{K}^{-} - \Pi_{K}^{+} \right) \right ) F_2\left(W^2,\tilde{Q}^2-\frac{Q^2}{4}\right) }{\left( \left( P + \tilde{q} \right)^2 - W^2 + i \varepsilon \right) \left( \left( P - \tilde{q} \right)^2 - W^2 + i \varepsilon \right)}, 
\eer
with 
\ber 
A & = & \frac{8 e^2}{M^2}  \frac{\left(1-\varepsilon\right)G_E}{\varepsilon G_E^2 + \tau G^2_M}, \qquad \qquad B =  \frac{\tilde{Q}^2+\frac{Q^2}{4}}{\tilde{Q}^2-\frac{Q^2}{4}}\left( Q^2 \left( P\cdot\tilde{q} \right) - 4 \left(K\cdot P\right) \left( K\cdot\tilde{q} \right)  \right), \nonumber \\
C & = & 2 \left(K\cdot P\right) \left( \left(K\cdot P\right)^2 - \frac{Q^2}{4}  P^2 \right) +  \frac{Q^2}{\tilde{Q}^2-\frac{Q^2}{4}}  \left(P\cdot\tilde{q} \right)  \left(\left(K\cdot\tilde{q} \right) P^2 - \left(P\cdot\tilde{q} \right) \left(K\cdot P\right) \right) , \nonumber\\
D & = &  \left( P^2 + \frac{ \left( P \cdot \tilde{q} \right)^2 Q^2}{\left(\tilde{Q}^2-\frac{Q^2}{4}\right)^2 }\right) \left( \left(K\cdot P\right) \left( K\cdot\tilde{q} \right)  - \frac{Q^2}{4} \left( P\cdot\tilde{q} \right)  \right)  \nonumber \\
& & - 2 \left(P\cdot\tilde{q} \right) \frac{ \tilde{Q}^2-\frac{3 Q^2}{4} }{ \tilde{Q}^2-\frac{Q^2}{4}  } \left( \left(K\cdot P\right)^2  - \frac{Q^2}{4}  P^2  \right). \label{abc}
\eer

\section{$ \mathrm{T}_1 $ subtraction function contribution}
\label{sec5}

Before discussing the inelastic contribution to $ \delta_{2 \gamma} $, we first discuss the subtraction term in the $ \mathrm{T}_1 $ amplitude regardless of the {\it near-forward} approximation of Eq. (\ref{vvcs_agreement_with_elastic}). The subtraction function $ \mathrm{T}_1 \left(0, Q^2\right) $  is expressed in terms of only two VVCS amplitudes $ B_1 $ and $ B_3 $ as
\ber \label{forward_T1}
\mathrm{T}_1 \left(0, Q^2\right) = \alpha \left( B_1 \left(Q^2\right) Q^2 + B_3\left(Q^2\right) Q^4 \right),
\eer
with the relevant VVCS tensor structures given by
\ber
T^{\mu \nu}_1 & = & - \left(q_1\cdot q_2\right) g^{\mu \nu} + q_1^\mu q^\nu_2,  \\
T^{\mu \nu}_3 & = & q_1^2 q_2^2 g^{\mu \nu} + \left(q_1\cdot q_2\right)  q_1^\nu q^\mu_2- \frac{q^2_1 + q^2_2}{2} \left( q_1^\nu q^\mu_1 + q_2^\nu q^\mu_2 \right)+ \frac{q^2_1 - q^2_2}{2} \left( q_1^\nu q^\mu_1 - q_2^\nu q^\mu_2 \right).
\eer
The TPE correction due to the subtraction term, $ \delta^{subt}_{2 \gamma}  $, arising from the amplitudes $ B_1, ~B_3 $, is obtained from Eq. (\ref{TPE_correction}) as
\ber \label{TPE_subtraction_Q_4}
\delta^{subt}_{2 \gamma} & \sim & \mathop{\mathlarger{\int}} \frac{ i \mathrm{d}^4 \tilde{q}}{\left( 2 \pi \right)^4} \left( B_1 \Pi_{Q}^{+} \Pi_{Q}^{-}  \left( \tilde{Q}^2 +  \frac{Q^2}{4}\right) + B_3 \right)   \Pi_{K}^{+} \Pi_{K}^{-} \nonumber \\
&&  \times \left \{   \frac{Q^2}{4} \left(P\cdot\tilde{q} \right)  \left(K\cdot\tilde{q} \right)   - \left(K \cdot P \right) \left(K\cdot\tilde{q} \right)^2 \right \} \equiv I_1 + I_2.
\eer
We express the loop integral of the first term of Eq. (\ref{TPE_subtraction_Q_4}), denoted by $ I_1 $, as
\ber \label{subtraction_1_initial}
I_1 & = &   \frac{Q^2}{4} \int \frac{ i \mathrm{d}^4 \tilde{q}}{\left( 2 \pi \right)^4}  \left( B_1  \Pi_{Q}^{+} \Pi_{Q}^{-}  \left( \tilde{Q}^2 +  \frac{Q^2}{4}\right) + B_3 \right)   \left(P\cdot\tilde{q} \right)  \left(K\cdot\tilde{q} \right) \Pi_{K}^{+} \Pi_{K}^{-} 
\nonumber \\
& = & \frac{Q^2}{16} \int \frac{ i \mathrm{d}^4 \tilde{q}}{\left( 2 \pi \right)^4}  \left( B_1  \Pi_{Q}^{+} \Pi_{Q}^{-}  \left( \tilde{Q}^2 +  \frac{Q^2}{4}\right) + B_3 \right)   \left(P\cdot\tilde{q} \right)  \left( \Pi_{K}^{-} - \Pi_{K}^{+} \right) =  a_K \left(K \cdot P \right),
\eer
where we have defined $ a_K $ as
\ber
 a_K K^\mu = \frac{Q^2}{16} \int \frac{ i \mathrm{d}^4 \tilde{q}}{\left( 2 \pi \right)^4}  \left( B_1  \Pi_{Q}^{+} \Pi_{Q}^{-}  \left( \tilde{Q}^2 +  \frac{Q^2}{4}\right) + B_3 \right) \tilde{q}^\mu  \left( \Pi_{K}^{-} - \Pi_{K}^{+} \right).
\eer
On the other hand, since $ K^2 = Q^2/4 $, we can rewrite the integral $ I_1 $ as
\ber \label{subtraction_12_initial}
I_1 & = & \left(K \cdot P \right) \int \frac{  i \mathrm{d}^4 \tilde{q}}{\left( 2 \pi \right)^4} \left( B_1  \Pi_{Q}^{+} \Pi_{Q}^{-}  \left( \tilde{Q}^2 +  \frac{Q^2}{4}\right) + B_3 \right) \left(K\cdot\tilde{q} \right)^2 \Pi_{K}^{+} \Pi_{K}^{-},
\eer
which exactly cancels the integral $ I_2 $ from the second term in Eq. (\ref{TPE_subtraction_Q_4}). Consequently, the subtraction function contribution to unpolarized elastic electron-proton scattering vanishes in the limit of massless electrons.

It is instructive to study the subtraction function contribution to the TPE amplitude $ T^{2 \gamma} $ of elastic lepton-proton scattering, accounting for a finite lepton mass. The amplitude is given by
\ber
& &  T^{2 \gamma} =  - e^2  \mathop{\mathlarger{\int}} \frac{i \mathrm{d}^4 \tilde{q}}{\left( 2 \pi \right)^3} \frac{ L^{\mu \nu }  \bar{N}(p',s') M_{\mu \nu} N(p,s)}{\left(\left(\tilde{q} - \frac{q}{2} \right)^2 - \mu^2\right)\left(\left(\tilde{q} + \frac{q}{2} \right)^2 - \mu^2\right)} ,
\eer
with the leptonic tensor
\ber
& & \hspace{0.2cm} L^{\mu \nu }  = \bar{u}\left(k',h'\right)  \left( \gamma^\mu \frac{\hat{K}-\hat{\tilde{q}}+m}{\left(K-\tilde{q}\right)^2-m^2} \gamma^\nu + \gamma^\nu \frac{\hat{K}+\hat{\tilde{q}}+m}{\left(K+\tilde{q}\right)^2-m^2} \gamma^\mu  \right) u\left(k,h\right). \nonumber 
\eer

We study the TPE contribution due to the VVCS amplitude $ B_1 $  first. The $  - \left(q_1\cdot q_2\right) g^{\mu \nu } $ term contribution is given by
\ber \label{gmunuterm}
e^4 \mathop{\mathlarger{\int}} \frac{ i \mathrm{d}^4 \tilde{q}}{\left( 2 \pi \right)^4}  B_1 \left( \bar{N} N \right) \cdot \left( 2 m \left( \tilde{q}^2 + \frac{Q^2}{4} \right) \bar{u} u + 4 \left( K \cdot \tilde{q} \right)  \bar{u} \hat{\tilde{q}}u  \right)  \left( \tilde{q}^2 + \frac{Q^2}{4} \right) \Pi_{K}^{+} \Pi_{K}^{-}\Pi_{Q}^{+} \Pi_{Q}^{-} .
\eer
Using the gauge symmetry in the tensor $ T_1^{\mu \nu} $ the $ q_1^\mu q^\nu_2 $ term contribution is given by
\ber \label{expansion}
-  e^4 Q^2 \mathop{\mathlarger{\int}} \frac{ i \mathrm{d}^4 \tilde{q}}{\left( 2 \pi \right)^4}  B_1   \left( \bar{N} N \right) \cdot \left( 2 \left(K \cdot \tilde{q} \right)  \bar{u} \hat{\tilde{q}}u  \right) \Pi_{K}^{+} \Pi_{K}^{-}\Pi_{Q}^{+} \Pi_{Q}^{-}.
\eer
Denoting
\ber \label{expansion}
-  e^4 Q^2 \mathop{\mathlarger{\int}} \frac{ i \mathrm{d}^4 \tilde{q}}{\left( 2 \pi \right)^4}  B_1  \cdot \left( 2 \left(K \cdot \tilde{q} \right) \tilde{q^\mu}  \right) \Pi_{K}^{+} \Pi_{K}^{-}\Pi_{Q}^{+} \Pi_{Q}^{-} = a_K^{\left(1\right)} K^\mu ,
\eer
the contribution of the $ q_1^\mu q^\nu_2 $ term is given by
\ber
m a_K^{(1)} \left( \bar{N} N \right) \cdot \left(\bar{u} u\right).
\eer
Note that, according to the symmetry properties of the VVCS amplitudes \cite{Drechsel:1997xv}, the expansion of Eq. (\ref{expansion}) also contains no $ q^\mu $ term for the case of VVCS amplitude $ B_1\left(q_1^2, q_2^2, \left(q_1 \cdot q_2\right), \left(P \cdot \tilde{q}\right)\right) $ subtracted at an arbitrary point $ \left(P \cdot \tilde{q}\right) = M \nu_0 $. The amplitude $ B_3 $ contributes through only one tensor structure $ q_1^2 q_2^2 g^{\mu \nu} $ in the similar way as the amplitude $ B_1 $ contributes through the structure $  - \left(q_1\cdot q_2 \right) g^{\mu \nu } $. 

We conclude that the $ \mathrm{T}_1 $ subtraction function contributes to the $ \left(\bar{u} u\right) \cdot \left( \bar{N} N \right)  $ term in the elastic lepton-proton scattering amplitude. This contribution is suppressed by the lepton mass \footnote{That is equivalent to a subtraction in the dispersion relation for the $ \bar F_4 $ invariant amplitude, in the notation of Ref. \cite{Gorchtein:2004ac}.}. In the language of effective field theories, the chiral (or axial) symmetry on the lepton side forbids the $ \left(\bar{u} u\right) \cdot \left( \bar{N} N \right)  $  type structure for massless electrons, and therefore the contribution of the subtraction function vanishes in the massless lepton limit.

\section{Evaluation of the inelastic TPE contribution}
\label{sec6}

In this section, we describe the strategy to evaluate the inelastic TPE correction $ \delta^{inel}_{2 \gamma} $, which we will express as a two-dimensional integral over the unpolarized proton structure functions $ F_1 $ and $ F_2 $:
\ber \label{TPE_expression}
\delta^{inel}_{2 \gamma} = \mathop{\mathlarger{\int}} \mathrm{d} W^2    \mathrm{d} \tilde{Q}^2 && \left \{ w_1\left(W^2,\tilde{Q}^2\right) F_1\left(W^2, \tilde{Q^2} - \frac{Q^2}{4}\right) \right. \nonumber \\
&& \left. + w_2\left(W^2,\tilde{Q}^2\right) F_2 \left(W^2, \tilde{Q^2} - \frac{Q^2}{4}\right) \right \}. \nonumber \\
\eer

To obtain the result of Eq. (\ref{TPE_expression}), starting from Eqs. (\ref{TPE_inelastic1}) and (\ref{TPE_inelastic2}), we will perform the integration, using two choices of Breit frames with the aim to cross check the method and its application. In a first choice, which we denote as $P$-frame, the kinematics of the external particles is defined by
\ber \label{pframe}
 K = \left(K_0,0,0,|\vec{K}|\right), \qquad  P = P\left(1,0,0,0\right),\qquad  q = Q\left(0,1,0,0\right).
\eer
In a second choice, which we denote as $K$-frame, the kinematics of the external particles is defined by
\ber \label{kframe}
 K = K\left(1,0,0,0\right), \qquad  P = \left(P_0,0,0,|\vec{P}|\right),\qquad  q = Q\left(0,1,0,0\right).
\eer

\begin{figure}[h]
\begin{center}
\includegraphics[width=.75\textwidth]{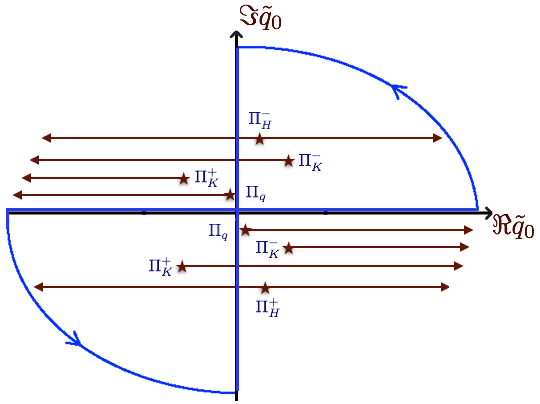}
\end{center}
\caption{The position of the $ \tilde{q}_0 $ poles for different propagators. The lepton poles ($ \Pi_K$) and hadronic poles ($ \Pi_H $) contribute only in a limited region of integration variables.}
\label{poles_position}
\end{figure}

In either of these frames, we evaluate the weighting functions $ w_1, ~w_2 $ by performing a Wick rotation  in Eqs. (\ref{TPE_inelastic1}) and (\ref{TPE_inelastic2}) in the complex $ \tilde{q}_0 $ variable plane. The resulting TPE correction is given by the sum of the integral along the imaginary $ \tilde{q}_0 $ axis and the contributions from the poles which are crossed by the integration contour.  In addition to the photon and lepton propagator poles, there exist the hadronic poles coming from the dispersion relation propagators:
\ber
 \Pi_H^{\pm} = \frac{ 1}{\left( \tilde{q} \pm P \right)^2 - W^2}.
\eer
The pole positions are shown in Fig. \ref{poles_position}.

\begin{figure}[h]
\begin{center}
\includegraphics[width=.75\textwidth]{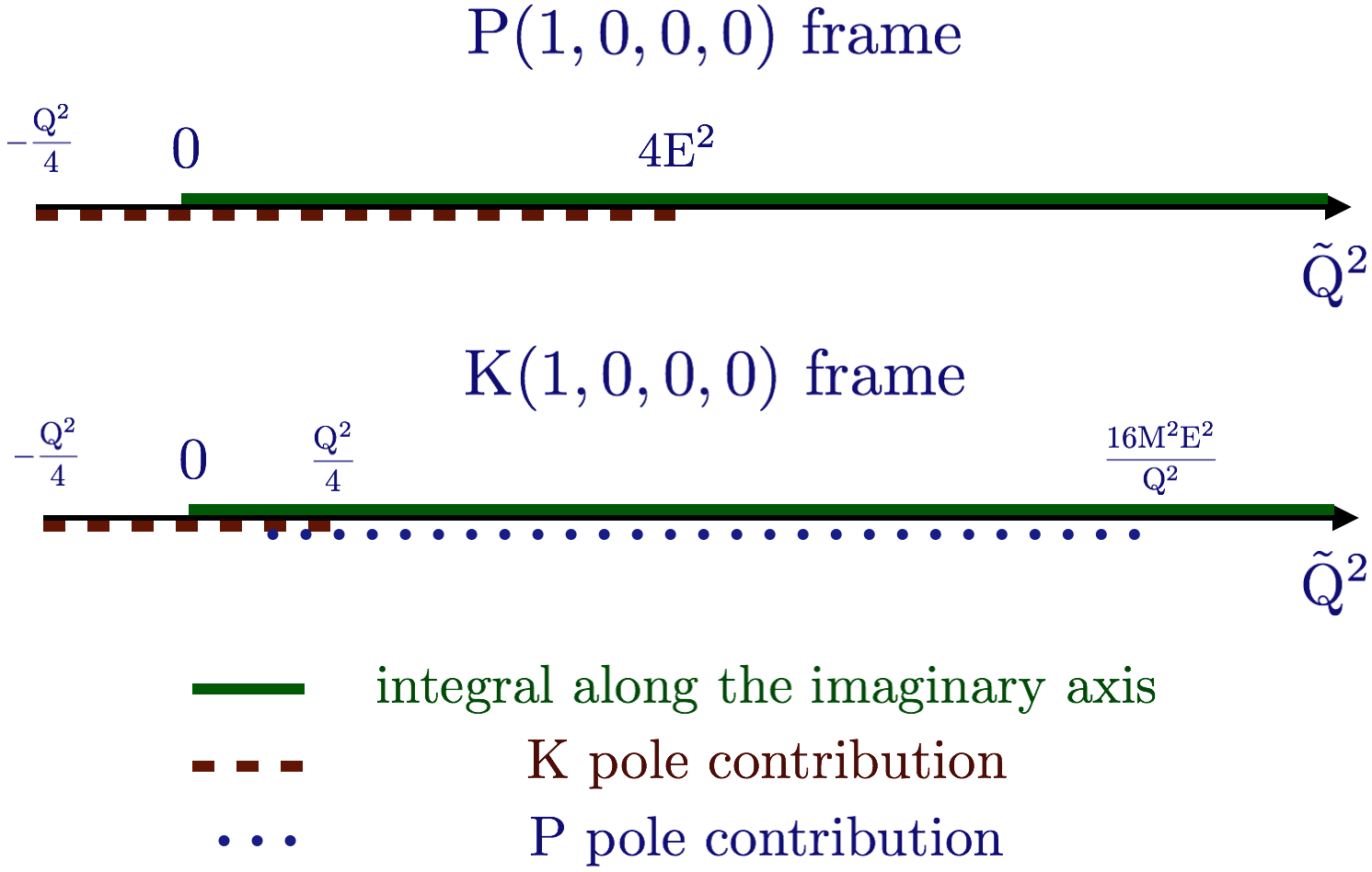}
\end{center}
\caption{Integration ranges for the different contributions to $ \delta^{inel}_{2 \gamma}$.}
\label{poles_position_Q2_line}
\end{figure}

The integration contour does not cross the photon poles $ \Pi_q $. The expressions of Eqs. (\ref{TPE_inelastic1}) and (\ref{TPE_inelastic2}) are symmetric with respect to the change of integration variable $ \tilde{q} \to - \tilde{q}$. Exploiting this symmetry, we only need to calculate the residues of the upper half plane poles and double the result. The leptonic pole $ \Pi^-_K $ and the hadronic pole $ \Pi^-_H $ are moving poles; they contribute in the limited region of $ W^2,  ~\tilde{Q}^2 $ variables. We show the corresponding regions in Fig. \ref{poles_position_Q2_line} for the limit of low-$Q^2$ scattering.

We provide the expressions for the regions of these poles contributions and weighting functions $ w_1, w_2$ in Appendix \ref{weights_expressions}.

\section{Leading $ Q^2 \ln Q^2 $ inelastic TPE contribution}
\label{sec7}

In this section, we describe the way to obtain the leading $ Q^2 \ln Q^2 $ inelastic TPE contribution in terms of the total photo absorption cross section $ \sigma_T $  \cite{Brown:1970te} in our approach. For the region of small momentum transfer, we exploit the approximation \cite{Drechsel:2002ar}:
\ber \label{structure_functions_approximation}
\label{photoabsorption1} F_1 \left(W^2,  \tilde{Q}^2 \right) & \approx & \frac{ \left(P \cdot \tilde{q} \right)}{ \pi e^2 } \sigma_T \left(W^2\right), \\
\label{photoabsorption2} F_2 \left(W^2,  \tilde{Q}^2 \right) & \approx & \frac{ \tilde{Q}^2}{ \pi e^2 } \sigma_T\left(W^2\right).
\eer

In the $ P$-frame, the leading correction is coming from the leptonic pole, whereas in the $K$-frame, it originates from the hadronic pole. We present its derivation in the $K$-frame. The pole $ \tilde{q}_0 = P_0 - \sqrt{ \left( \vec{P}- \vec{\tilde{q}} \right)^2 + W^2 } $ contributes in the invariant mass region $ W^2 < P_0^2 \approx 4 \left( K \cdot P \right)^2/Q^2$. For the leading contribution, we consider the $ W^2 $ integration up to infinity. For the case of small momentum transfer values, the integration region in the $ \tilde{Q}^2 $ variable is given by
\ber
 \alpha_0  Q^2  \le  \tilde{Q}^2 \le \frac{16 \left( K \cdot P \right)^2}{Q^2},
\eer
with 
\ber
\alpha_0 = \left( \frac{W^2-M^2}{4 \left( K \cdot P \right)} \right)^2.
\eer
We can replace the upper integration limit by  infinity, but the approximation of Eqs. (\ref{photoabsorption1}) and (\ref{photoabsorption2}) in terms of the photoabsorption cross section is valid only up to some hadronic scale $ \Lambda $, that reproduces the $ \tilde{Q}^2 $ behavior of unpolarized structure functions~\footnote{Note that the numerical evaluation of Ref. \cite{Gorchtein:2014hla} corresponds with a specific choice of hadronic scale, $ \Lambda^2 = \left(W^2-M^2\right)^2/\left(M^2 + 2 M E \right)$.}. In the present work, we will go beyond such approximation and directly use the unpolarized proton structure functions $ F_1, ~F_2 $ with their $ \tilde{Q}^2 $ dependence as an input. The integration region $ \tilde{Q}^2 > \Lambda^2 $ will not contribute to the term proportional to $ Q^2 \ln Q^2 $, and therefore we integrate up to the squared hadronic scale $ \Lambda^2 $.

The logarithmic term comes from the region $ \tilde{Q}^2 \gg Q^2 $. Accounting also for the pole condition $ \left( P \cdot \tilde{q} \right) = \left(P^2 - \tilde{Q}^2 - W^2 \right)/2 \gg Q^2 $, we get
\ber
\delta_{2 \gamma}^{F1} & = &  \frac{2 Q^2 e^2}{\left(K\cdot P\right) }   \Re \mathop{\mathlarger{\int}} \limits^{~~ \infty}_{W^2_{thr}} \mathrm{d} W^2   \mathop{\mathlarger{\int}} \frac{\mathrm{d}^3 \vec{\tilde{q}}}{\left( 2 \pi \right)^3}  \frac{K \tilde{q}_0   \left( \Pi_{K}^{-} - \Pi_{K}^{+} \right) }{\tilde{Q}^4  } \frac{ F_1\left(W^2,\tilde{Q}^2\right) }{ P_0 - \tilde{q}_0 }, \\
\delta_{2 \gamma}^{F2} & = &  4 Q^2 e^2 \Re \mathop{\mathlarger{\int}} \limits^{~~ \infty}_{W^2_{thr}} \mathrm{d} W^2  \mathop{\mathlarger{\int}}\frac{\mathrm{d}^3 \vec{\tilde{q}}}{\left( 2 \pi \right)^3} \frac{    \left(P\cdot\tilde{q} \right)  \left( \Pi_{K}^{-} - \Pi_{K}^{+} \right) - \left(K\cdot P\right) \left( \Pi_{K}^{-} + \Pi_{K}^{+} \right)}{ \tilde{Q}^4 \left( W^2 - P^2 + \tilde{Q}^2  \right)}\frac{ F_2\left(W^2,\tilde{Q}^2\right) }{ P_0 - \tilde{q}_0 }. \nonumber \\
\eer

With account of the proton structure functions approximation of Eqs. (\ref{photoabsorption1}) and (\ref{photoabsorption2}) the leading term of the TPE correction expansion is expressed as
\ber
\delta_{2 \gamma}^{F1} & = &  - \frac{Q^4}{\left(K\cdot P\right)^2 }  \Re \mathop{\mathlarger{\int}} \limits^{~~ \infty}_{W^2_{thr}} \frac{\mathrm{d} W^2}{16 \pi^3}  \sigma_T\left(W^2 \right)  \mathop{\mathlarger{\int}} \limits^{~~\Lambda^2}_{\alpha_0 Q^2} \mathrm{d} \tilde{Q}^2 \frac{ \left(P \cdot \tilde{q} \right)}{\tilde{Q}^4}   \left( L_1^{-} - L_1^{+} \right), \\
\delta_{2 \gamma}^{F2} & = &  \frac{Q^2 K}{\left(K\cdot P\right) }  \Re \mathop{\mathlarger{\int}} \limits^{~~ \infty}_{W^2_{thr}} \frac{\mathrm{d} W^2}{4 \pi^3}  \sigma_T\left(W^2 \right) \mathop{\mathlarger{\int}} \limits^{~~\Lambda^2}_{\alpha_0 Q^2} \mathrm{d} \tilde{Q}^2 \frac{ \left(K\cdot P\right) \left( L_0^- + L_0^+ \right) -  \left(P\cdot\tilde{q} \right) \left( L_0^- - L_0^+ \right) }{\tilde{Q}^2 \left(P \cdot \tilde{q} \right)}, 
\eer
with the following integrals:
\ber
L^\pm_0 & = & \mathop{\mathlarger{\int}} \limits^{~~ \tilde{q}_0^M}_{0} \mathrm{d} \tilde{q}_0 \Pi_{K}^\pm = - \mathop{\mathlarger{\int}} \limits^{~~ \tilde{q}_0^M}_{0}  \frac{\mathrm{d} \tilde{q}_0}{\tilde{Q}^2 \mp Q \tilde{q}_0} = \pm \frac{1}{Q} \ln \frac{W^2 -P^2 \mp 2 \left(K\cdot P\right) }{W^2 -P^2}, \\
L^\pm_1 & = & \mathop{\mathlarger{\int}} \limits^{~~ \tilde{q}_0^M}_{0} \tilde{q}_0 \mathrm{d} \tilde{q}_0 \Pi_{K}^\pm = - \mathop{\mathlarger{\int}} \limits^{~~ \tilde{q}_0^M}_{0}  \frac{\tilde{q}_0 \mathrm{d} \tilde{q}_0}{\tilde{Q}^2 \mp Q \tilde{q}_0} \approx \frac{\tilde{Q}^2}{Q^2} \left( \ln  \frac{W^2 -P^2 \mp 2 \left(K\cdot P\right) }{W^2 -P^2} \pm \frac{2 \left( K \cdot P \right)}{W^2-P^2} \right), \nonumber \\
\tilde{q}_0^M & = & \frac{2 \left( K \cdot P \right)}{W^2-P^2} \frac{\tilde{Q}^2}{Q}.
\eer

Performing the $ \tilde{Q}^2 $ integration, we obtain the leading logarithmic contributions,
\ber \label{brown_result}
\label{brown1} \delta_{2 \gamma}^{F1} & = &  \frac{Q^2}{8 \pi^3} \ln \left( \frac{Q^2}{\Lambda^2} \right) \int \limits_{W^2_{thr}}^{\infty} \frac{ \mathrm{d} W^2} {M E} \left( \frac{W^2 - P^2}{4 M E} \ln \frac{2 \left( K \cdot P \right) + P^2 - W^2 }{2 \left( K \cdot P \right) - P^2 + W^2 } + 1 \right) \sigma_T(W^2) , \\
\label{brown2} \delta_{2 \gamma}^{F2} & = & \frac{Q^2}{8 \pi^3} \ln \left( \frac{Q^2}{\Lambda^2} \right)  \int \limits_{W^2_{thr}}^{\infty} \frac{ \mathrm{d} W^2} {M E} \ln \frac{(W^2 - P^2)^2}{\left( 2 \left( K \cdot P \right) + P^2 - W^2 \right) \left(2 \left( K \cdot P \right) - P^2 + W^2 \right)} \sigma_T(W^2)    \nonumber \\ 
& + & \frac{Q^2}{4 \pi^3} \ln \left( \frac{Q^2}{\Lambda^2} \right) \int \limits_{W^2_{thr}}^{\infty} \frac{ \mathrm{d} W^2} {W^2 - P^2} \ln \frac{2 \left( K \cdot P \right) + P^2 - W^2 }{2 \left( K \cdot P \right) - P^2 + W^2 }  \sigma_T(W^2) , 
\eer
which sum up to the known expression of Ref. \cite{Brown:1970te}. 

When also accounting for the $ Q^2 $ terms in the expansion of Eq. (\ref{TPE_expression}) for the TPE correction, the hadronic scale $ \Lambda $ dependence in Eqs. (\ref{brown1}) and (\ref{brown2}) drops out.

\section{Results and discussion}
\label{sec8}

Having verified the leading $ Q^2 \ln Q^2 $ contribution for the inelastic TPE, we next discuss the numerical evaluation of Eq. (\ref{TPE_expression}) including the $ \tilde{Q}^2 $ dependence of the structure functions. 

\begin{figure}[h]
\begin{center}
\includegraphics[width=1.\textwidth]{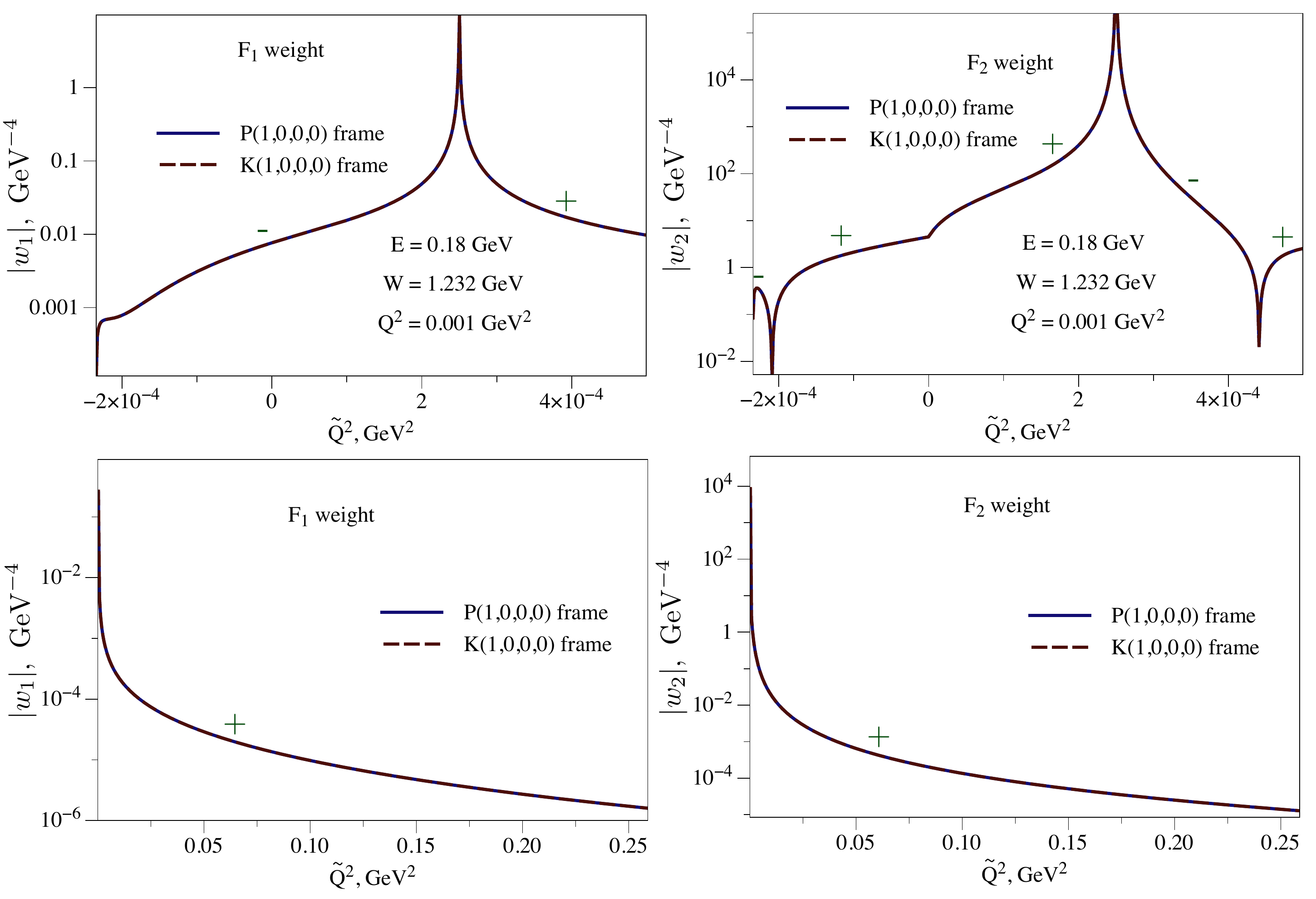}
\end{center}
\caption{The weighting functions of Eq. (\ref{TPE_expression}), with the absolute value plotted and where the sign labels show the corresponding signs of $ w_1 $ and $ w_2 $ for $ e^- p $ scattering. As a check, the weighting functions are evaluated in two frames, yielding exactly the same result. The kinematics is chosen as shown on the figure. In the upper panels, the weighting functions $ w_1 $ and $ w_2 $ are shown in the region $ -\frac{Q^2}{4} \le \tilde{Q}^2 \le \frac{Q^2}{2} $ and in the lower panels for $  \frac{Q^2}{2} \le \tilde{Q}^2 \le 8 E^2 $.}
\label{electron_two_frames}
\end{figure}

As a numerical check, we evaluate the weighting functions $ w_1 $ and $ w_2 $ both in the $ K$-frame and $ P$-frame, in Figs. \ref{electron_two_frames}. When adding the integral along the imaginary axis with the pole contributions, we checked that we obtain the same result in both frames. Despite the finite $ \tilde{Q}^2 $ ranges of the poles contribution, which are distinct in the $ K$- and $ P$-frames, the resulting weighting functions are continuous. The boundaries of the $K$-pole region and the lower boundary of the $ P$-pole region, shown in Fig. \ref{poles_position_Q2_line}, are covered by Figs. \ref{electron_two_frames}. The weighting functions $ w_1 $ and $ w_2 $ have a singularity at $ \tilde{Q}^2 = Q^2 / 4 $, when the photons are on their mass shell in our approximation, i. e. $ q_1^2 = q_2^2 = 0 $. The weighting function $ w_2 $ has a discontinuity in the first derivative at $ \tilde{Q}^2 = 0 $, where the integral along the imaginary axis starts to contribute.

\begin{figure}[h]
\begin{center}
\includegraphics[width=0.8\textwidth]{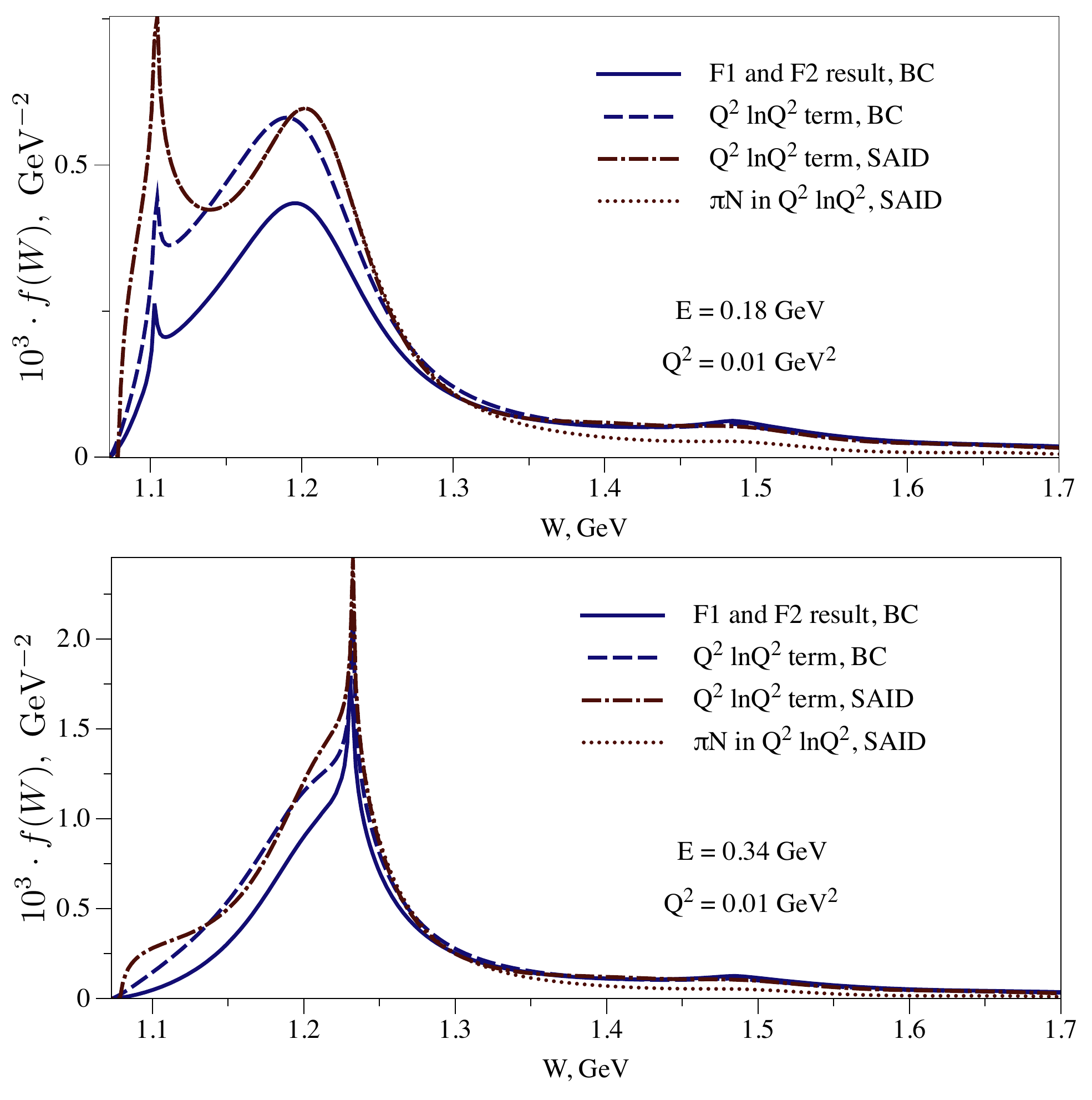}
\end{center}
\caption{$ W^2 $ integrand $ f $ of Eq. (\ref{integrand_TPE}) for two different external kinematics. The integrand with the unpolarized proton structure functions from BC \cite{Christy:2007ve} is shown by the blue solid curve. The leading logarithmic correction in the approximation of Eqs. (\ref{brown1}) and (\ref{brown2}) with $ \Lambda = 0.6 ~\mathrm{GeV}$ and the photoabsorption cross section from the fit of BC (SAID \cite{Arndt:2002xv}) is shown by the blue dashed (the red dash-dotted) curves. The dominant $\pi N$-channel contribution is shown for the SAID fit by the red dotted curve.}
\label{W_integrand}
\end{figure}

We perform the $ \tilde{Q}^2 $ integration first and express the resulting TPE correction $ \delta_{2 \gamma}^{inel} $ in terms of the $ W^2 $ integral as
\ber \label{integrand_TPE}
\delta_{2 \gamma}^{inel} = \mathop{\mathlarger{\int}} \limits^{~~\infty}_{W^2_{thr}} f \left(W\right) \mathrm{d} W^2.
\eer
In Fig. \ref{W_integrand}, we present the result for the $ W^2 $ integrand for different inputs of proton structure functions and compare our full calculation with the approximation of Eqs. (\ref{brown1}) and (\ref{brown2}) with $ \Lambda \approx 0.6 ~\mathrm{GeV}$. To describe the $ \tilde{Q}^2 $ dependence of the unpolarized structure functions, we use the empirical fit performed by Christy and Bosted (BC) \cite{Christy:2007ve}, while for the logarithmic approximation in the Fig. \ref{W_integrand}, we also use the SAID Partial-Wave Analysis Facility \cite{Arndt:2002xv} for the photoabsorption cross section. The BC fit is valid in the region $ 0 < \tilde{Q}^2 < 8 ~\mathrm{GeV}^2 $, $ M + m_\pi < W < 3.1 ~\mathrm{GeV} $. This fit is not very accurate in the threshold region near $ \tilde{Q}^2 = 0 $, although it is still compatible with the error bars of the photoproduction cross sections. We checked that difference between using the SAID and BC fit in the region $ W < 1.15 ~\mathrm{GeV} $ and $ 0.9 \times Q^2 /4 < \tilde{Q}^2 < 1.1 \times Q^2/4 $ only amounts to the relative change in $ \delta_{2 \gamma}^{inel} $ at the $ 3\%$-$4 \% $ level for the kinematics shown in Fig. \ref{W_integrand}. For the result of $ \delta_{2 \gamma}^{inel} $ beyond the $ Q^2 \ln Q^2 $ approximation, we use the BC fit for the region $  0< \tilde{Q}^2 < 12 ~\mathrm{GeV}^2 $, where the integrand behaves in a smooth way. The relative contribution to $ \delta_{2 \gamma }^{inel} $ from the region $  8< \tilde{Q}^2 < 12 ~\mathrm{GeV}^2 $ is smaller than $ 0.1 \%$. We perform the $ W^2 $ integration up to $ W = 4 ~\mathrm{GeV} $. The extrapolation from $ W = 3.1 ~\mathrm{GeV} $ (upper bound of the BC fit) to $ W = 4 ~\mathrm{GeV} $  leads to an additional relative contribution to $ \delta_{2 \gamma }^{inel} $ of less than $ 2\%$. We have also checked on the SAID parametrization that the region $ W > 4 ~\mathrm{GeV} $ has a relative contribution to $ \delta_{2 \gamma }^{inel} $ of less than $ 2\%$-$3\%$, when interpolating the SAID parametrization to the Regge behavior. The main inelastic TPE contribution is given by the $ \pi N $-channel. The singular peak at $ W^2 = M^2 + 2 M E $ corresponds to the quasireal photon singularity (when both photons in the two-photon box are quasireal and collinear with either lepton). This singularity appears only for the beam energies above the pion threshold.

\begin{figure}[h]
\begin{center}
\includegraphics[width=1.0\textwidth]{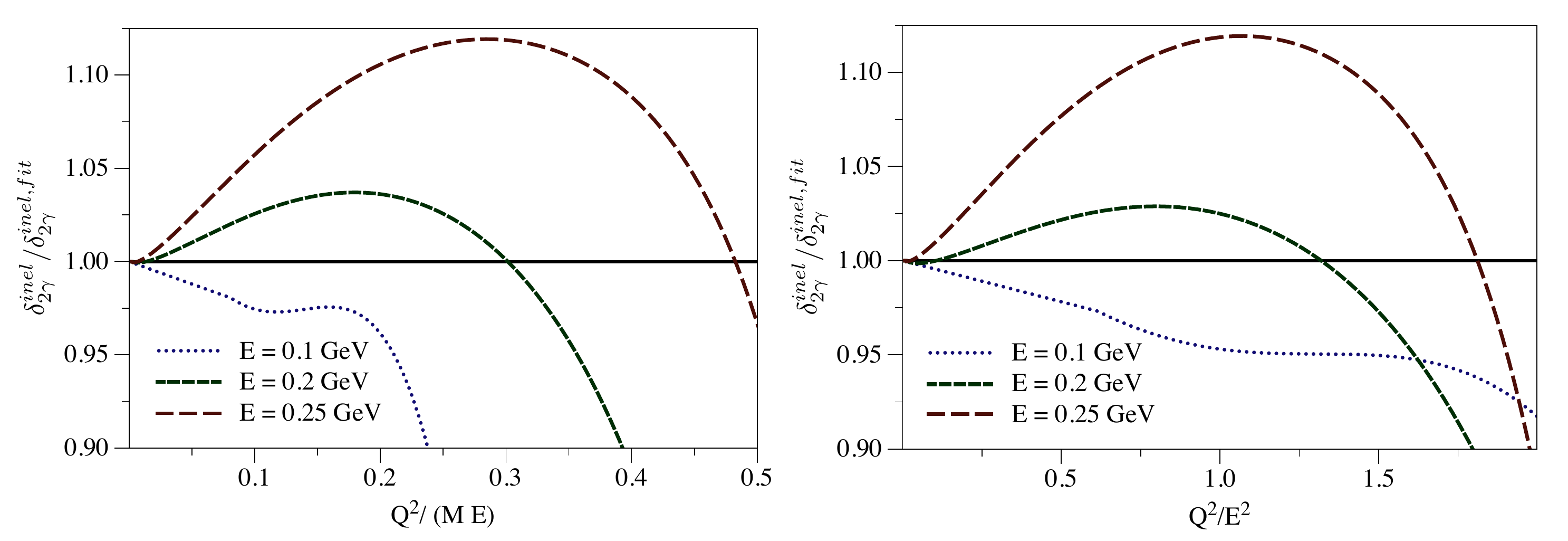}
\end{center}
\caption{$ Q^2$ dependence of ratio of the inelastic TPE correction $ \delta_{2 \gamma}^{inel} $ to the low-$ Q^2 $ fit of the form $ \delta_{2 \gamma}^{inel,fit} =  a\left(E\right) Q^2 \ln Q^2 +  b\left(E\right)  Q^2 $. The ratio is shown as function of $ Q^2/\left( M E \right) $ on the left panel and $ Q^2 / E^2 $ on the right panel.}
\label{F1F2_expansion}
\end{figure}

In order to clarify the validity of the inelastic TPE estimates, we study numerically the low-$Q^2$ expansion of the $ \delta_{2 \gamma}^{inel} $ coming from the $ F_1 $ and $ F_2 $ structure functions. In Fig. \ref{F1F2_expansion}, we present the ratio between the TPE correction $ \delta_{2 \gamma}^{inel} \left( Q^2 \right) $ and the low-$ Q^2 $ fit $ \delta_{2 \gamma}^{inel, fit} \left( Q^2 \right) $ in the form of Eq. (\ref{inelastic_expansion}) for the energies of available data. We compare the $ Q^2 / E^2 $ and $ Q^2 / \left( M E \right) $ expansions. We perform the fit in either the variable $ Q^2 / E^2 $ or $ Q^2 / \left( M E \right) $ in a range which is 100 times smaller than the range displayed in Fig. \ref{F1F2_expansion}. The comparison of our full calculation with such a fit over an extended range provides us then with a quantitative argument on the $Q^2$ range where such an expansion holds. If we use as a criterion that the full calculation stays within $ 10\% $ of the fit, we can see from Fig. \ref{F1F2_expansion} that for energies corresponding with available data an expansion of the type of Eq. (\ref{inelastic_expansion}) holds for $ Q^2 \lesssim E^2 $ and requires $ Q^2 \lesssim \left( M E \right) / 5 $. We expect the same type of expansion for the TPE contributions from other amplitudes.

\begin{figure}[h]
\begin{center}
\includegraphics[width=.7\textwidth]{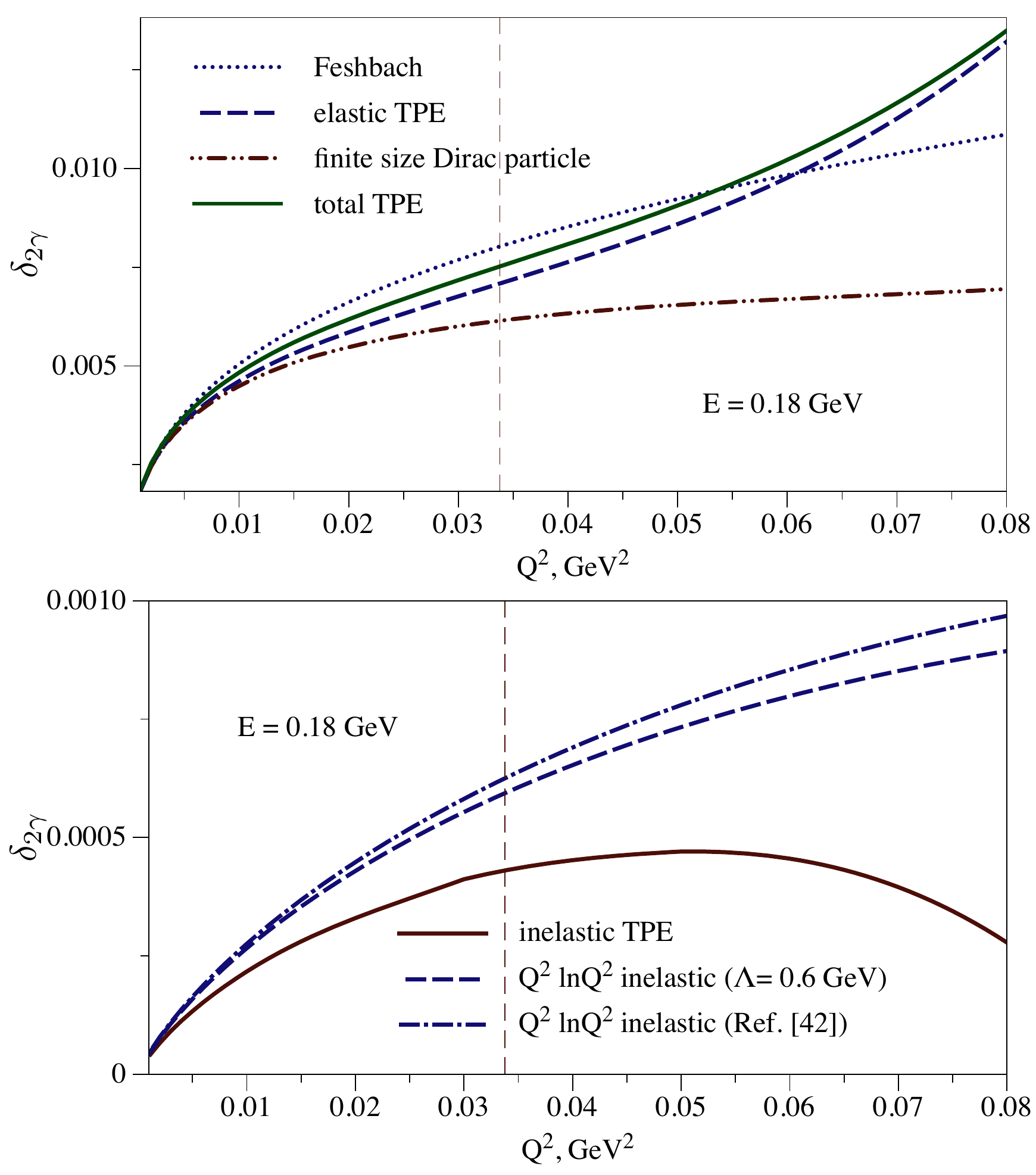}
\end{center}
\caption{$ Q^2$ dependence of the TPE correction $ \delta_{2 \gamma} $ to $ e^{-} p \to e^{-} p $ for lab electron energy $ E = 0.180 ~\mathrm{GeV} $ (for which the kinematically allowed region is $ Q^2 < 0.094 ~\mathrm{GeV}^2$). The Feshbach term for pointlike particles, the elastic TPE contribution based on the box graph evaluation with dipole form factors \cite{Tomalak:2014sva}, and the total TPE correction as the sum of elastic TPE and inelastic TPE are presented (upper panel). The inelastic contribution is compared with the leading logarithmic approximation of Eqs. (\ref{brown1}) and (\ref{brown2}) with $ \Lambda = 0.6 ~\mathrm{GeV} $  and the result of Ref. \cite{Gorchtein:2014hla} (lower panel). The experimental input for the proton structure functions is taken from the Christy-Bosted fit \cite{Christy:2007ve}. The vertical dashed lines restrict the region of validity of the expansion for the inelastic term $ Q^2 \lesssim \left( M E \right) / 5 $ as follows from Fig. \ref{F1F2_expansion}.}
\label{electron_results}
\end{figure}

The resulting inelastic TPE correction as a function of $ Q^2 $ for the beam energy $ E = 0.18~\mathrm{GeV} $ is shown in Fig. \ref{electron_results}. We compare the Feshbach term for pointlike particles, the elastic TPE correction in the model with dipole form factors \cite{Tomalak:2014sva}, the approximation of Eqs. (\ref{brown1}) and (\ref{brown2})  with $ \Lambda \approx 0.6 ~\mathrm{GeV}$ and the result of Ref. \cite{Gorchtein:2014hla}. The difference between both $ Q^2 \ln Q^2 $ curves comes from the term of order $ Q^2 $ and appears due to the different choices of $ \Lambda $. The approximation of Eqs. (\ref{brown1}) and (\ref{brown2}) implies the same hadronic scale for all intermediate states, while the hadronic scale in Ref. \cite{Gorchtein:2014hla} depends on the intermediate state. We present our results for the total TPE correction including the $ Q^2 $ dependence in the structure functions and extrapolate them to the region $ Q^2 \gtrsim M E $. Increasing the momentum transfer the inelastic TPE correction shows a clear departure from the $ Q^2 \ln Q^2 $ term, reducing the latter value. With account of the inelastic intermediate states the TPE correction in the low-$Q^2$ region comes closer to the Feshbach correction in comparison with the elastic TPE correction only. The inelastic TPE correction has the same order of magnitude and the opposite sign in comparison with the proton form factor effects in the elastic TPE.

\begin{figure}[h]
\begin{center}
\includegraphics[width=.8\textwidth]{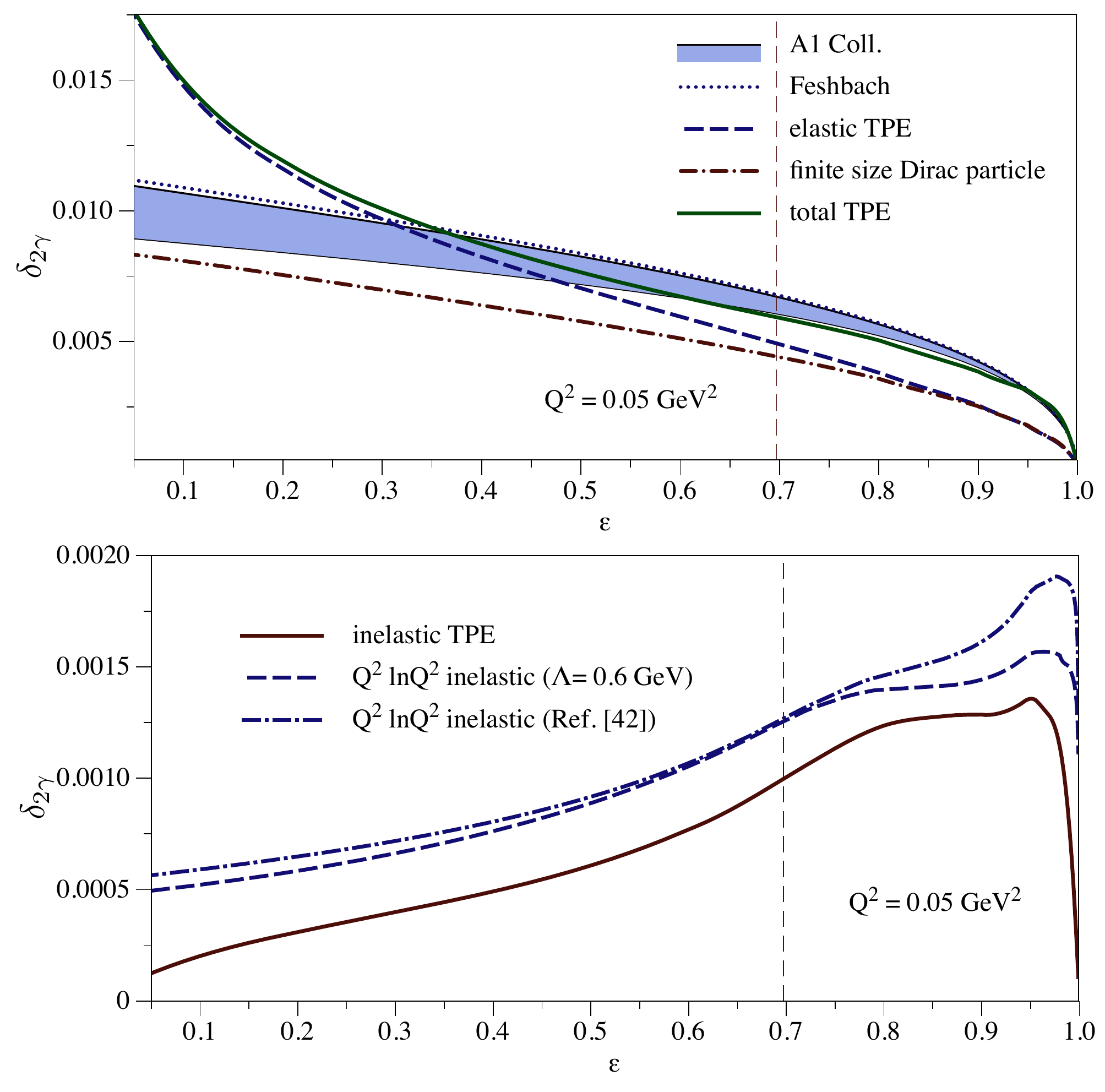}
\end{center}
\caption{Same as Fig. \ref{electron_results}, but for the fixed value $ Q^2 = 0.05 ~\mathrm{GeV}^2 $ as function of $ \varepsilon $ in comparison with the empirical TPE fit using the data of Ref. \cite{Bernauer:2013tpr} (A1 Collaboration, blue bands).}
\label{electron_results2}
\end{figure}

\begin{figure}[h]
\begin{center}
\includegraphics[width=.8\textwidth]{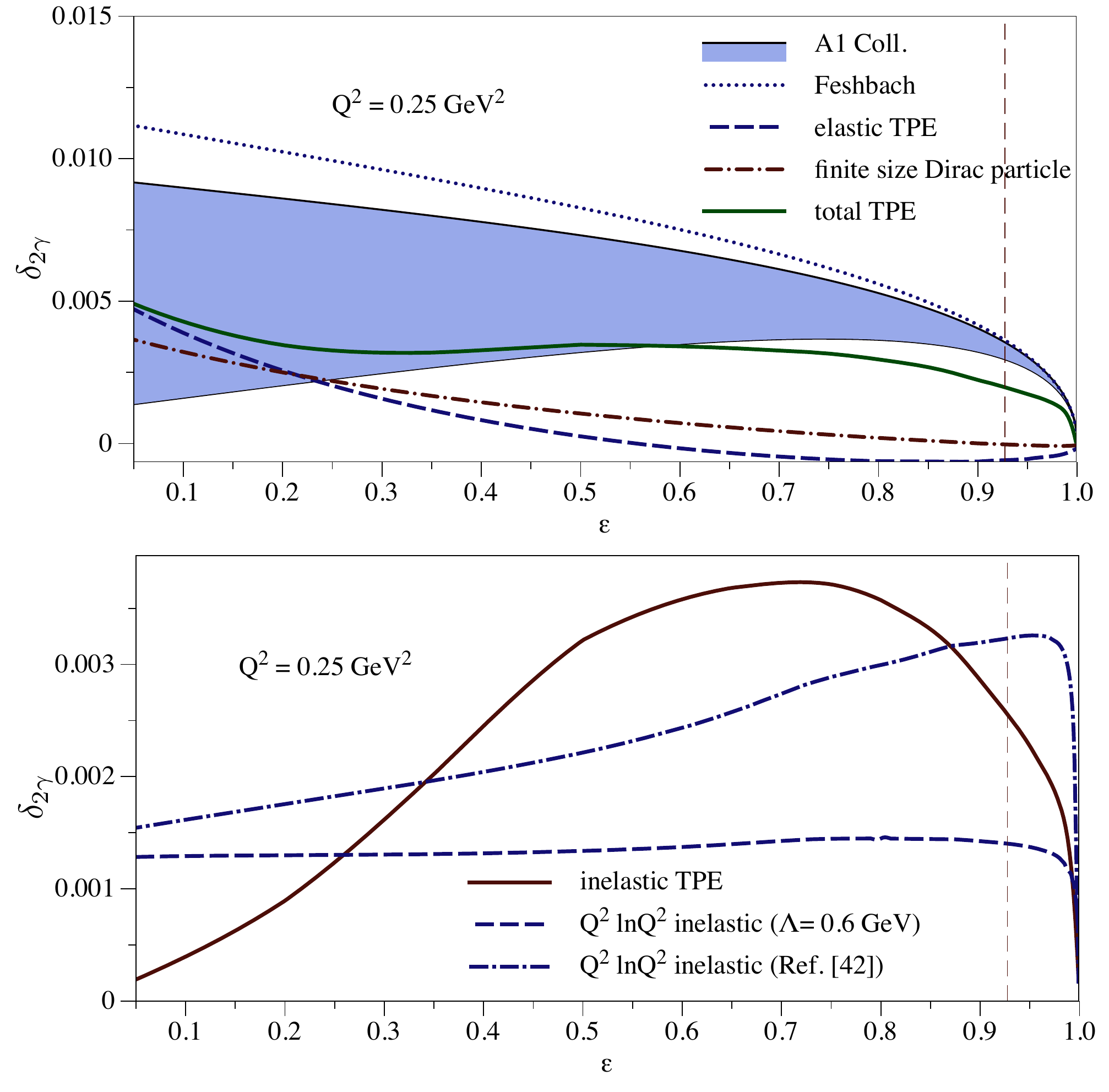}
\end{center}
\caption{Same as Fig. \ref{electron_results2}, but for the fixed value $ Q^2 = 0.25 ~\mathrm{GeV}^2 $.}
\label{electron_results3}
\end{figure}

We compare the TPE corrections as a function of the $ \varepsilon $ variable for $ Q^2 = 0.05 ~\mathrm{GeV}^2 $ ($ Q^2 = 0.25 ~\mathrm{GeV}^2 $) in Fig. \ref{electron_results2} (\ref{electron_results3}). The inelastic proton excitations compensate the proton form factor effects, and the resulting TPE correction comes closer to the Feshbach term. For the small momentum transfer $ Q^2 = 0.05 ~\mathrm{GeV}^2 $, where we expect the validity of the {\it near-forward} approximation for $ \varepsilon \gtrsim 0.7 $ or $ Q^2 / \left( M E \right) \lesssim 1/5 $, our calculation is in good agreement with the empirical TPE fit of Ref. \cite{Bernauer:2013tpr} in the region $ \varepsilon > 0.35-0.4 $ as one notices from Fig. \ref{electron_results2}. One should definitely account for contributions beyond two unpolarized proton structure functions for smaller $ \varepsilon $ values. We show the region of small $ \varepsilon $ with the aim to illustrate the characteristic features of our calculation. Increasing the momentum transfer, as shown on Fig. \ref{electron_results3}, the predicted TPE correction is found to be in reasonable agreement with the empirical fit of Ref. \cite{Bernauer:2013tpr}, confirming the proton charge radius values extracted with this TPE correction.

\section{Conclusions and outlook}
\label{sec9} 
In this work, we have evaluated the leading terms in the momentum transfer expansion due to the inelastic TPE contribution to the unpolarized electron-proton scattering. We expressed this inelastic TPE contribution as an integral over the VVCS amplitudes. The VVCS tensor was then approximated by the two unpolarized forward Compton scattering amplitudes. The latter are obtained as a dispersion integral over the unpolarized proton structure functions. This approximation allows one to reproduce the existent results for the low momentum transfer expansion of the TPE correction and goes beyond the $ Q^2 \ln Q^2 $ term in such an expansion. Our results for the kinematics of the MAMI experiments show that the inelastic contribution has the same sign as the Feshbach correction. With account of the inelastic contribution, the TPE correction comes closer to the Feshbach term for pointlike particles in comparison with the elastic contribution only. In the limit of low $Q^2$, the TPE correction is in good agreement with the empirical TPE fit of the A1 Collaboration at MAMI \cite{Bernauer:2013tpr} at small momentum transfers. This agreement confirms the value of the extracted electric charge radius from the data of Ref. \cite{Bernauer:2013tpr}. Increasing the momentum transfer, the TPE correction starts to deviate from the data fit, still following its shape. To further test the region of applicability of the {\it near-forward} approximation, a detailed dispersion relation analysis of separate channels contributions (e.g. the $ \pi N$-channel) for nonforward kinematics will be needed. Such a comparative study will be performed in a forthcoming work.

\appendix

\section{Some VVCS amplitudes and tensor structures}
\label{elastic_F1F1_structures}

The three nonvanishing VVCS amplitudes for the case of two Dirac couplings $ \gamma^\mu $ in the photon-proton-proton vertices in Fig. \ref{VVCS} are given by
\ber
B_2^{point} & = & - \frac{1}{M} \Pi_{P}^{-} \Pi_{P}^{+},  \nonumber \\
B_{10}^{point} & = & \frac{1}{2 M} \Pi_{P}^{-} \Pi_{P}^{+}, \nonumber \\
B_{17}^{point} & = & \frac{\left(P\cdot \tilde{q}\right)}{M} \Pi_{P}^{-} \Pi_{P}^{+}, 
\eer
where the relevant VVCS tensor structures have the following form:
\ber
T^{\mu \nu}_2 & = & - 4 \left(P\cdot \tilde{q}\right)^2 g^{\mu \nu} - 4 \left(q_1\cdot q_2\right) P^\mu P^\nu + 4 \left(P\cdot\tilde{q} \right) \left(P^{\nu} q^{\mu}_1 + P^{\mu} q^{\nu}_2\right), \nonumber \\
T^{\mu \nu}_{10} & = & - 8 \left(q_1\cdot q_2\right) P^\mu P^\nu + 4 \left(P\cdot \tilde{q}\right) \left(P^{\nu} q^{\mu}_1 + P^{\mu} q^{\nu}_2\right) + 4 M \left(q_1\cdot q_2\right) \left(P^{\mu} \gamma^\nu + P^\nu \gamma^\mu\right) \nonumber \\
& & - 4 M \left(P\cdot \tilde{q}\right) \left(q_1^{\mu} \gamma^\nu + q_2^\nu \gamma^\mu\right)  - 2 \left(P\cdot \tilde{q}\right) \left \{ \left( q_2^{\nu} \gamma^\mu - q_1^\mu \gamma^\nu \right) \gamma.\tilde{q} - \gamma.\tilde{q} \left( q_2^{\nu} \gamma^\mu - q_1^\mu \gamma^\nu \right) \right \} \nonumber \\
& & - 2  \left(q_1\cdot q_2\right) \left(P\cdot \tilde{q}\right) \left( \gamma^\nu \gamma^\mu - \gamma^{\mu} \gamma^\nu\right) + M \left(q_1\cdot q_2\right)   \left\{ \left( \gamma^\nu \gamma^\mu - \gamma^{\mu} \gamma^\nu \right) \gamma.\tilde{q} + \gamma.\tilde{q} \left( \gamma^\nu \gamma^\mu - \gamma^{\mu} \gamma^\nu \right) \right\}, \nonumber \\ 
T^{\mu \nu}_{17} & = & - 4  \left(P \cdot \tilde{q}\right) g^{\mu \nu} + 2 \left(P^{\nu} q^{\mu}_1 + P^{\mu} q^{\nu}_2\right) + 4M g^{\mu \nu }  \gamma.\tilde{q}- 2 M  \left(q_1^{\mu} \gamma^\nu + q_2^\nu \gamma^\mu\right)  \nonumber \\
& &   + \left\{  \left( q_2^{\nu} \gamma^\mu - q_1^\mu \gamma^\nu\right) \gamma.\tilde{q} - \gamma.\tilde{q}\left( q_2^{\nu} \gamma^\mu - q_1^\mu \gamma^\nu\right) \right\}  +    \left(q_1\cdot q_2\right) \left( \gamma^\nu \gamma^\mu - \gamma^{\mu} \gamma^\nu \right).  
\eer

\section{Evaluation of the forward limit of the Feshbach correction}
\label{feshbach_and_IR_term}

The leading term in the low momentum transfer ($Q$) expansion for $ \delta_{2 \gamma} $ comes from the following term in Eq. (\ref{lp_elastic_integral}):
\ber
\delta_{2 \gamma} \to - \frac{16 M Q^2 e^2}{E} \mathop{\mathlarger{\int}} \frac{  i \mathrm{d}^4 \tilde{q}}{\left( 2 \pi \right)^4} \Pi_{P}^{+} \Pi_{P}^{-}\Pi_{K}^{+} \Pi_{K}^{-}\Pi_{Q}^{+} \Pi_{Q}^{-}   \left(\tilde{q}^2 + \frac{Q^2}{4}\right) \left(K\cdot\tilde{q}\right)^2.
\eer

We perform the integration in Euclidean space in this Appendix. The Euclidean coordinates  $ \tilde{q}_E $ can be expressed through the Minkowski coordinates $ \left(\tilde{\nu}, \vec{\tilde{q}}\right)$ by
\ber
 \tilde{q}_E \left(\tilde{q}_{E}^0,\vec{\tilde{q}}_E\right) = \tilde{q}_E \left(- i \tilde{\nu},\vec{\tilde{q}}\right), \qquad \quad  \left(a_E\cdot b_E\right) = - \left(a\cdot b\right),
\eer
where we use the index "E" for the notation of vectors in Euclidean space. 

The leading term of the Feshbach correction can then be cast into the form
\ber \label{feshbach_expression}
\delta^{0}_{2 \gamma} = \frac{M Q^2 e^2}{E}  \int \frac{  \mathrm{d}^4 \tilde{q}_E}{\left( 2 \pi \right)^4}  \left(\Pi_{P}^{-} + \Pi_{P}^{+}\right)\left( \Pi_{Q}^{+} + \Pi_{Q}^{-} \right)\Pi_{Q},
\eer
with $ \Pi_Q = 1/\left(\tilde{q}_E^2+\frac{Q^2}{4}+\mu^2\right) $. We can expand $ \Pi^{\pm}_{P,Q} $ as Gegenbauer polynomials [$ C_n\left(z\right) $] generating functions,
\ber \label{propagators}
\frac{1}{\left(\tilde{q}_E + \frac{q_E}{2} \right)^2 +  \mu^2} & = & \frac{ 2 z_\mu}{\tilde{q}_E Q} \sum \limits^{\infty}_{n=0} \left(-1\right)^n z_\mu^n C_n\left(\hat{\tilde{\vec{q}}}_E \hat{\vec{q}}_E\right), \nonumber \\
\frac{1}{\left(\tilde{q}_E - \frac{q_E}{2} \right)^2 +  \mu^2}  & = & \frac{ 2 z_\mu}{\tilde{q}_E Q} \sum \limits^{\infty}_{n=0} z_\mu^n C_n\left(\hat{\tilde{\vec{q}}}_E \hat{\vec{q}}_E\right), \nonumber \\
 \frac{1}{\left(\tilde{q}_E + P_E\right)^2 + M^2} & = & \frac{ i z_M}{\tilde{q}_E P_E} \sum \limits^{\infty}_{n=0} \left(-1\right)^n \left(i z_M\right)^n C_n\left(\hat{\tilde{\vec{q}}}_E \hat{\vec{P}}_E\right), \nonumber \\
 \frac{1}{\left(\tilde{q}_E - P_E\right)^2 + M^2} & = & \frac{ i z_M}{\tilde{q}_E P_E} \sum \limits^{\infty}_{n=0} \left(i z_M\right)^n C_n\left(\hat{\tilde{\vec{q}}}_E \hat{\vec{P}}_E\right), 
\eer
with
\ber
z_M & = & \frac{Q}{4Mx}\left( 1-x^2 + \sqrt{ \left(1-x^2 \right)^2 + \frac{16 M^2}{Q^2} x^2} \right), \nonumber \\
z_\mu & = & \frac{1 + x^2 + \tilde{\mu}^2 - \sqrt{\left(1 + x^2 + \tilde{\mu}^2\right)^2 - 4 x^2}}{2x},  \nonumber \\
x & = & \frac{2 \tilde{q}_E}{Q}, \qquad \tilde{\mu}^2 = 4\mu^2/Q^2.
\eer
Using the Gegenbauer polynomials value $ C_n\left(0\right) = \left(-1\right)^{n/2} \cdot \left(1+\left(-1\right)^n \right)/2 $ and the orthogonality relation for vectors $ q, x, y $ in Euclidean space
\ber \label{orthogonality}
\int \mathrm{d} \Omega \left(\hat{q}\right) C_n\left(\hat{q} \hat{x}\right) C_m\left(\hat{q} \hat{y}\right) = \frac{2 \pi^2}{n+1} \delta_{m,n} C_n\left(\hat{x} \hat{y}\right),
\eer
the integral of Eq. (\ref{feshbach_expression}) simplifies to the following expression:
\ber \label{sum_integral_Feshbach}
\delta^{0}_{2 \gamma} =  \frac{4\alpha}{\pi} \frac{Q^2}{E} \mathop{\mathlarger{\int}} \limits^{~~\infty}_0  \frac{ \mathrm{d} \tilde{q}_E }{\tilde{q}_E^2 + \frac{Q^2}{4}} \frac{\tilde{q}_E}{Q} \sum \limits^{\infty}_{n=0} \frac{\left( z_M z_\mu \right)^{2n+1} }{2n+1} = \frac{2\alpha}{\pi} \frac{Q}{E} \mathop{\mathlarger{\int}} \limits^{~~\infty}_0  \mathrm{d} x \frac{x}{1+x^2} \ln \left( \frac{1+z_M z_\mu}{1-z_M z_\mu} \right).
\eer
For $ x \to 0 $, the product $ z_M z_\mu \to \frac{Q}{2M}\left(1 -\frac{\tilde{\mu}^2}{1 + \tilde{\mu}^2}\right)$ is IR finite. Numerically, the result of the integral does not depend on small $ \frac{Q}{M} $ and $ \tilde{\mu}^2 $ values, so we can neglect terms of order $ \tilde{\mu}^2, ~\frac{Q}{M} $. The resulting TPE correction is given by
\ber
\delta^{0}_{2 \gamma} = \frac{2\alpha}{\pi} \frac{Q}{E} \mathop{\mathlarger{\int}} \limits^{~~\infty}_0  \mathrm{d} x \frac{x}{1+x^2} \ln \left( \frac{1+x}{|1-x|} \right) + O \left(Q^2 \ln Q^2 \right) =  \frac{\alpha \pi  Q}{2E} + O \left( Q^2 \ln Q^2 \right),
\eer
reproducing the low-$ Q^2 $ behavior of the Feshbach correction; see Eq. (\ref{deltaIR_Feshbach_corr}).

\section{TPE correction in terms of the unpolarized proton structure functions}
\label{weights_expressions}

In this Appendix, we present the expressions for the pole contributions as well as the contribution from the integral along the imaginary axis to the weighting functions $ w_1,~w_2 $, which appear in the TPE correction of Eq. (\ref{TPE_expression}).

We first present the results in the $P$-frame, defined by Eq. (\ref{pframe}).
The contribution to the weighting functions $ w_1,~w_2 $ arising from the leptonic pole $ \tilde{q}_0 = K_0 - \sqrt{  \left( \vec{K} - \vec{\tilde{q}} \right)^2 + m^2 }$ in the $P$-frame is given by
\ber
w_1 \left(W^2, \tilde{Q}^2\right) & = &  \frac{2 \alpha}{\pi}  \frac{G_E}{\varepsilon G_E^2 + \tau G^2_M} \frac{P^2}{M^2}\frac{Q^2 }{\left(K\cdot P\right)^2 + M^4 \tau \left( 1 + \tau \right)} \frac{1}{ |\vec{K}|}  \frac{1}{\tilde{Q}^2 + \frac{Q^2}{4} } \frac{ P^2 \tilde{q}_0^2 \tilde{B} }{ \left( W^2 - P^2 + \tilde{Q}^2 \right)^2 } , \nonumber \\
w_2 \left(W^2, \tilde{Q}^2\right) & = &  \frac{2\alpha}{\pi}  \frac{G_E}{\varepsilon G_E^2 + \tau G^2_M}\frac{P^2}{M^2}  \frac{Q^2}{\left(K\cdot P\right)^2 + M^4 \tau \left( 1 + \tau \right) } \frac{1}{ |\vec{K}|}   \frac{1}{\tilde{Q}^2 + \frac{Q^2}{4} } \frac{ \tilde{C} + \tilde{D} }{ W^2 - P^2 + \tilde{Q}^2 } , \nonumber \\
\eer
with coefficients $ \tilde{B} $ and $ \tilde{C} + \tilde{D} $ following from Eqs. (\ref{abc}) as
\ber \label{notation_K_pole_P_frame}
\tilde{B} & = & -\left( \tilde{Q}^2 + \frac{Q^2}{4} \right) \left(K\cdot P\right) - 2 \frac{\tilde{Q}^2+\frac{Q^2}{4}}{\tilde{Q}^2-\frac{Q^2}{4}} \frac{Q^2}{4} P \tilde{q_0} , \nonumber \\
\tilde{C} + \tilde{D} & = & - \left(K\cdot P\right) \left( \left(K\cdot P\right)^2 - \frac{3 Q^2}{16} P^2 - \frac{ \tilde{Q}^2}{4} P^2 \right) + \frac{Q^2}{8} P^3 \tilde{q}_0   + \frac{Q^2}{8} \frac{Q^2 P^3 \tilde{q}_0}{\tilde{Q}^2 - \frac{Q^2}{4}} \nonumber \\
& &  + P \tilde{q}_0 \frac{ \tilde{Q}^2-\frac{3 Q^2}{4} }{ \tilde{Q}^2-\frac{Q^2}{4}  } \left(K\cdot P\right)^2 + \frac{3}{4} \frac{ P^2 \tilde{q}_0^2 Q^2}{\tilde{Q}^2-\frac{Q^2}{4} } \left(K\cdot P\right) + \frac{Q^2}{8} \frac{ P^3 \tilde{q}_0^3 Q^2}{\left(\tilde{Q}^2-\frac{Q^2}{4}\right)^2 }.
\eer
In Eq. (\ref{notation_K_pole_P_frame}), $ \tilde{q}_0^n $ stands for the sum of two integrals with either $ \pm $ signs:
\ber
 \tilde{q}_0^n \to \int \limits^{\tilde{q}_0^+}_{0 ~\mathrm{or} ~\tilde{q}_0^-}  \frac{ \mathrm{d} \tilde{q}_0 }{\sqrt{\left(\tilde{Q}^2 - \frac{Q^2}{4} \right)^2 -  Q^2 \tilde{q}_0^2 + \frac{Q^2}{4 | \vec{K}|^2} \left(  \tilde{Q}^2 - \frac{Q^2}{4} + 2 \tilde{q}_0  K_0 \right)^2 }} \frac{ \tilde{q}_0^n }{ P^2 \pm 2 P \tilde{q}_0 - \tilde{Q}^2  - W^2}. \nonumber \\
\eer
The integration regions are given by
\ber \label{Q2_regions}
0 \le \tilde{q}_0 \le \tilde{q}_0^{+} && \quad \mathrm{for} \quad \left( -|\vec{K}| + \sqrt{  K_0^2 - m^2 } \right)^2 \le \tilde{Q}^2 \le \left( |\vec{K}| + \sqrt{  K_0^2 - m^2 } \right)^2 , \nonumber \\
 \tilde{q}_0^{-} \le \tilde{q}_0 \le \tilde{q}_0^{+}  && \quad \mathrm{for} \quad - \left( K - m \right)^2 \le \tilde{Q}^2 \le \left( -|\vec{K}| + \sqrt{  K_0^2 - m^2 } \right)^2 ,
\eer
with
\ber
\tilde{q}_0^{\pm} = \frac{K_0}{2} \left( 1 - \frac{m^2}{K^2} \right) &-& \frac{K_0 \tilde{Q}^2}{2 K^2} \pm \frac{|\vec{K}|}{2 K^2} \sqrt{ \left( \left(K+m\right)^2 + \tilde{Q}^2 \right) \left( \left(K-m\right)^2 + \tilde{Q}^2 \right)}.
\eer

We do not consider the hadronic pole in the $P$-frame. It contributes only in the region of large momentum transfer:
\ber
 Q^2 \ge 8 M m_\pi \left( 1 + \frac{m_\pi}{2M} \right) \approx 1.09~\mathrm{GeV}^2.
\eer

The contribution to the weighting functions arising from the integral along the imaginary axis in the $P$-frame is given by
\ber
w_1 \left(W^2, \tilde{Q}^2\right) & = &   \frac{2 \alpha}{\pi} \frac{G_E}{\varepsilon G_E^2 + \tau G^2_M}  \frac{P^2}{M^2} \frac{Q^2}{\left(K \cdot P \right)^2 + M^4 \tau \left( 1 + \tau \right)} \frac{  \tilde{Q}^2 }{\tilde{Q}^2 + \frac{Q^2}{4} } \nonumber \\
 & & \times \Re \mathop{\mathlarger{\int}} \limits^{~~\pi}_{0} \frac{\left(P\cdot\tilde{q} \right)^2}{ W^2 - P^2 + \tilde{Q}^2 }  \frac{  \tilde{B} \sin^2 \psi  \mathrm{d}  \psi}{\left( \left( P + \tilde{q} \right)^2 - W^2  \right) \left( \left( P - \tilde{q} \right)^2 - W^2 \right)}, \nonumber  \\
w_2 \left(W^2, \tilde{Q}^2\right) & = &   \frac{2 \alpha}{\pi} \frac{G_E}{\varepsilon G_E^2 + \tau G^2_M} \frac{P^2}{M^2} \frac{Q^2}{\left(K \cdot P \right)^2 + M^4 \tau \left( 1 + \tau \right)} \frac{  \tilde{Q}^2 }{\tilde{Q}^2 + \frac{Q^2}{4} }  \nonumber \\
 & & \times \Re \mathop{\mathlarger{\int}} \limits^{~~\pi}_{0}  \frac{ \left(  \tilde{C} + \tilde{D} \right ) \sin^2 \psi  \mathrm{d}  \psi}{ \left( \left( P + \tilde{q} \right)^2 - W^2  \right) \left( \left( P - \tilde{q} \right)^2 - W^2 \right)},
\eer
with the notations
\ber
\tilde{B} & = & - 2 \frac{\tilde{Q}^2+\frac{Q^2}{4}}{\tilde{Q}^2-\frac{Q^2}{4}} \left(K\cdot P\right) J_0 +  2 (  \tilde{Q}^2+ \frac{Q^2}{4})  \left(K\cdot P\right) I_0^c +  \frac{\tilde{Q}^2+\frac{Q^2}{4}}{\tilde{Q}^2-\frac{Q^2}{4}} Q^2 \left( P\cdot\tilde{q} \right) I_0^c, \nonumber \\
\tilde{C} + \tilde{D} & = & \left(  \left( P^2 + \frac{ \left(P\cdot\tilde{q} \right)^2 Q^2}{\left(\tilde{Q}^2-\frac{Q^2}{4}\right)^2 }\right) \left(K\cdot P\right)  + \frac{Q^2}{\tilde{Q}^2-\frac{Q^2}{4}}  \left(P\cdot\tilde{q} \right)   P^2 \right) \frac{J_0}{2}  \nonumber \\
 && + 2 \left(K\cdot P\right) \left( \left(K\cdot P\right)^2 - \frac{Q^2}{4} P^2 \right) I_0^c  - \frac{Q^2}{2}  \left(P\cdot\tilde{q} \right)  P^2 I_0^c - \frac{Q^2}{\tilde{Q}^2-\frac{Q^2}{4}}  \left(P\cdot\tilde{q} \right)^2 \left(K\cdot P\right) I_0^c  \nonumber \\
&& - \left(  P^2 + \frac{ \left(P\cdot\tilde{q} \right)^2 Q^2}{\left(\tilde{Q}^2-\frac{Q^2}{4}\right)^2 }\right) \frac{Q^2}{4} \left(P\cdot\tilde{q} \right) I_0^c - 2 \left(P\cdot\tilde{q} \right) \frac{ \tilde{Q}^2-\frac{3 Q^2}{4} }{ \tilde{Q}^2-\frac{Q^2}{4}  } \left( \left(K\cdot P\right)^2  - \frac{Q^2}{4} P^2 \right) I_0^c   \nonumber \\
&& - \left(  P^2 + \frac{ \left(P\cdot\tilde{q} \right)^2 Q^2}{\left(\tilde{Q}^2-\frac{Q^2}{4}\right)^2 }\right)  \left(K\cdot P\right) \left( \tilde{Q}^2-\frac{Q^2}{4} \right) \frac{I_0^c}{2},
\eer
and master integrals
\ber \label{integrals}
J_0 & = & \mathop{\mathlarger{\int}} \limits^{~~ 1}_{-1} \frac{\mathrm{d} x}{\sqrt{a^2+b^2 x^2}}   =  \frac{1}{b} \ln \frac{\sqrt{a^2+b^2}+b}{\sqrt{a^2+b^2} - b}, \nonumber \\
I_0^c & = & \mathop{\mathlarger{\int}} \limits^{~~ 1}_{-1} \frac{\mathrm{d} x}{\sqrt{a^2+b^2 x^2}} \frac{1}{c + g x}  =  \frac{1}{\sqrt{a^2 g^2 +c^2 b^2}} \ln \frac{ \left(\sqrt{a^2+b^2} c + \sqrt{a^2 g^2 + c^2 b^2} \right)^2 }{ a^2 \left(  c^2 - g^2 \right)} , \nonumber \\
a^2 &= & \left(\tilde{Q}^2 - \frac{Q^2}{4} \right)^2 -  Q^2 \tilde{q}_0^2 = \left(\tilde{Q}^2 + \frac{Q^2}{4} \right)^2 - b^2,  \qquad b^2 =  Q^2 \tilde{q}^2,  \qquad \tilde{q} = | \vec{\tilde{q}}|,
\eer
with
\ber
c = \tilde{Q}^2 -\frac{Q^2}{4} + 2 K_0 \tilde{q}_0 , \qquad g^2 =  4 \vec{K}^2 \tilde{Q}^2 \sin^2 \psi.
\eer
The integral along the imaginary axis contributes in the range $ 0 \le \tilde{Q}^2 \le \infty $.

As a check of our calculations, we also provide the expressions evaluated in the $K$-frame, defined by Eq. (\ref{kframe}). The leptonic pole $ \tilde{q}_0 = K - \sqrt{  \tilde{q}^2 + m^2 } $ contribution to $ w_1,~w_2 $ evaluated in the $K$-frame is given by
\ber
w_1 \left(W^2, \tilde{Q}^2\right) & = &  \frac{\alpha}{4 \pi} \frac{G_E}{\varepsilon G_E^2 + \tau G^2_M}  \frac{P^2}{M^2}  \frac{ \tilde{q} Q^3 }{\left(K\cdot P\right)^2 + M^4 \tau \left( 1 + \tau \right) } \frac{ 1 }{\tilde{Q}^2 - \frac{Q^2}{4}}, \nonumber \\
w_2 \left(W^2, \tilde{Q}^2\right) & = & \frac{\alpha}{4 \pi} \frac{G_E}{\varepsilon G_E^2 + \tau G^2_M}  \frac{P^2}{M^2}  \frac{ \tilde{q} Q^3}{\left(K\cdot P\right)^2 + M^4 \tau \left( 1 + \tau \right) }  \frac{ 1}{\tilde{Q}^2 + \frac{Q^2}{4}},
\eer
with the notations
\ber \label{notationsBCD}
\tilde{B} & = & c_+ I^{c_+}_0 - c_- I^{c_-}_0, \nonumber \\
\tilde{C} + \tilde{D} & = &   \frac{Q^2}{4 \left(\tilde{Q}^2-\frac{Q^2}{4}\right)^2 } c_0  \left( c_- I_0^{c_-} - c_+ I_0^{c_+} \right)  + \frac{16  \left( K \cdot P \right)^3 }{Q^2 c_0}\left( I_0^{c_-} + I_0^{c_+} \right)  \nonumber \\
&&   + \frac{2 \left( K \cdot P \right)  }{\tilde{Q}^2-\frac{Q^2}{4}}  \left( \left(  \frac{ \tilde{Q}^2-\frac{3 Q^2}{4} }{ Q^2  }4 \left( K \cdot P \right) + 2 P_0 \tilde{q}_0  \right) \left( I^{c_-}_0 - I^{c_+}_0 \right)  + 2 J_0 - c_+ I^{c_+}_0 - c_- I^{c_-}_0\right) \nonumber \\
&&   + \frac{ 4 K \tilde{q}_0 + \tilde{Q}^2 - \frac{5 Q^2}{4} }{\tilde{Q}^2-\frac{Q^2}{4}}  P^2  \left( I_0^{c_+} - I_0^{c_-} \right) + 2 \left( \tilde{q}_0 - 2 K \right)  \frac{P_0 P^2 }{c_0} \left( I_0^{c_+} + I_0^{c_-} \right),
\eer
in terms of the integrals of Eqs. (\ref{integrals}), where
\ber \label{notations}
c_{\pm}  =  W^2 - P^2 + \tilde{Q}^2 \mp 2 P^0 \tilde{q}_0, \qquad c_0 = W^2 - P^2 + \tilde{Q}^2, \qquad g^2  =   4  \frac{\left( K \cdot P \right)^2 - P^2 K^2 }{K^2}  \tilde{q}^2.~~~~~
\eer
The $ \tilde{Q}^2 $ integration region is given by
\ber
-\left(K-m\right)^2 \leq \tilde{Q}^2 \leq  \frac{Q^2}{4}.
\eer

The contribution to $ w_1,~w_2 $ arising from the hadronic pole $ \tilde{q}_0 = P_0 - \sqrt{\left( \vec{P} - \vec{\tilde{q}}\right)^2+W^2}$ in the $K$-frame is given by
\ber
w_1 \left(W^2, \tilde{Q}^2\right) & = &   \frac{\alpha}{2 \pi} \frac{G_E}{\varepsilon G_E^2 + \tau G^2_M}  \frac{P^2}{M^2} \frac{Q^2}{\left(K\cdot P\right)^2 + M^4 \tau \left( 1 + \tau \right) } \frac{1}{|\vec{P}|}  \frac{\tilde{B}}{\tilde{Q}^2 + \frac{Q^2}{4} }, \nonumber \\
w_2 \left(W^2, \tilde{Q}^2\right) & = & \frac{\alpha}{2 \pi} \frac{G_E}{\varepsilon G_E^2 + \tau G^2_M}  \frac{P^2}{M^2}  \frac{Q^2}{\left(K\cdot P\right)^2 + M^4 \tau \left( 1 + \tau \right) } \frac{1}{|\vec{P}|}    \frac{\tilde{C} + \tilde{D}}{\tilde{Q}^2 + \frac{Q^2}{4} }  \frac{ 1 }{\left(P \cdot \tilde{q} \right)},
\eer
with the notations
\ber
\tilde{B}  & = &  2 \frac{\tilde{Q}^2+\frac{Q^2}{4}}{\tilde{Q}^2-\frac{Q^2}{4}} \left( \left(K\cdot P\right) K \left( L_1^- - L_1^+ \right)  - \frac{Q^2}{4} \left(P\cdot\tilde{q} \right) \left( L_0^- - L_0^+ \right) \right) , \nonumber \\
\tilde{C} & = & \left(K\cdot P\right) \left( 2 \left(K\cdot P\right)^2 - \frac{Q^2 P^2}{2} - \frac{Q^2}{\tilde{Q}^2-\frac{Q^2}{4}}  \left(P\cdot\tilde{q} \right)^2  \right) \left( L_0^- + L_0^+ \right) \nonumber \\ 
&& + \frac{Q^2 K}{\tilde{Q}^2-\frac{Q^2}{4}}  \left(P\cdot\tilde{q} \right) P^2 \left( L_1^- + L_1^+ \right) , \nonumber\\
\tilde{D} & = &  \left( P^2 + \frac{ \left(P\cdot\tilde{q} \right)^2 Q^2}{\left(\tilde{Q}^2-\frac{Q^2}{4}\right)^2 }\right) \left( \left(K\cdot P\right) K \left( L_1^- - L_1^+ \right)  - \frac{Q^2}{4} \left(P\cdot\tilde{q} \right) \left( L_0^- - L_0^+ \right) \right) \nonumber \\
&&  - 2 \left(P\cdot\tilde{q} \right) \frac{ \tilde{Q}^2-\frac{3 Q^2}{4} }{ \tilde{Q}^2-\frac{Q^2}{4}} \left( \left(K\cdot P\right)^2 - \frac{Q^2 P^2}{4} \right)\left( L_0^- - L_0^+ \right). 
\eer
and master integrals
\ber
L^\pm_0 & = & \mathop{\mathlarger{\int}} \limits^{~~ \tilde{q}_0^M}_{0} \mathrm{d} \tilde{q}_0 \frac{\Pi_{K}^\pm}{D_\gamma}, \qquad L^\pm_1 = \mathop{\mathlarger{\int}} \limits^{~~ \tilde{q}_0^M}_{0} \tilde{q}_0 \mathrm{d} \tilde{q}_0 \frac{\Pi_{K}^\pm}{D_\gamma}, \nonumber \\
 D_\gamma & = & \sqrt{\left(\tilde{Q}^2 - \frac{Q^2}{4} \right)^2 -  Q^2 \tilde{q}_0^2 + \frac{Q^2}{ |\vec{P}|^2} \left( \left(P \cdot \tilde{q} \right) - P_0 \tilde{q}_0 \right)^2}, \nonumber \\
\tilde{q}_0^M & = & \frac{\left(P\cdot\tilde{q} \right) P_0 +|\vec{P}| \sqrt{\left(P\cdot\tilde{q} \right)^2 + P^2 \tilde{Q}^2 } }{P^2}.
\eer
The integration region is given by
\ber
\left(|\vec{P}| - \sqrt{P^2_0 - W^2}\right)^2 & \le & \tilde{Q}^2 \le \left(|\vec{P}| + \sqrt{P^2_0 - W^2}\right)^2, \nonumber \\
W^2_{thr} & \le & W^2 \le P^2_0.
\eer

The contribution to the weighting functions arising from the integral along the imaginary axis in the $K$-frame is given by
\ber
w_1 \left(W^2, \tilde{Q}^2\right)& = & \frac{\alpha}{ 8 \pi^2} \frac{G_E}{\varepsilon G_E^2 + \tau G^2_M} \frac{P^2}{M^2} \frac{Q^4}{\left( K \cdot P\right)^2 + M^4 \tau \left( 1 + \tau \right)}  \frac{\tilde{Q}^2}{\tilde{Q}^2 - \frac{Q^2}{4}} \nonumber \\
&& \times \Re \mathop{\mathlarger{\int}} \limits_{0}^{~~\pi} \frac{ \tilde{B} \sin^2 \psi \mathrm{d} \psi }{\tilde{Q}^2 - \frac{Q^2}{4} + 2 i K \tilde{Q} \cos \psi } , \nonumber \\
w_2 \left(W^2, \tilde{Q}^2\right)& = & \frac{\alpha}{ 8 \pi^2} \frac{G_E}{\varepsilon G_E^2 + \tau G^2_M} \frac{P^2}{M^2}  \frac{Q^4}{ \left( K \cdot P\right)^2 + M^4 \tau \left( 1 + \tau \right)} \frac{\tilde{Q}^2}{\tilde{Q}^2 + \frac{Q^2}{4} } \nonumber \\ 
&& \times \Re \mathop{\mathlarger{\int}} \limits_{0}^{~~\pi} \frac{ \left( \tilde{C} + \tilde{D} \right) \sin^2 \psi \mathrm{d} \psi }{\tilde{Q}^2 - \frac{Q^2}{4} + 2 i K \tilde{Q} \cos \psi}, \nonumber \\
\eer
with the notations of Eqs. (\ref{notationsBCD}), the integrals of Eqs. (\ref{integrals}), and notations of Eqs. (\ref{notations}). The integral along the imaginary axis contributes in the range $ 0 \le \tilde{Q}^2 \le \infty $.

In numerical evaluations of the TPE correction, we use the analytical expressions for the master integrals of the poles contributions. The integration over the hyperangle $ \psi $  for the integral along the imaginary axis is performed numerically.

The experimentally inaccessible region $ - Q^2 / 4 < \tilde{Q}^2 < 0   $, corresponding with negative $ \tilde{Q}^2 $ values, is present in the $ \tilde{Q}^2 $ integration. For this relatively small region, we approximate the unpolarized proton structure functions by the following relations, $ F_1\left(W^2, -\tilde{Q}^2 \right) \approx F_1\left(W^2, \tilde{Q}^2 \right) , ~F_2 \left(W^2, -\tilde{Q}^2 \right) \approx -F_2 \left(W^2, \tilde{Q}^2 \right)  $, according to the approximation of Eqs. (\ref{photoabsorption1}) and (\ref{photoabsorption2}).

\section*{Acknowledgements}

We thank Nikolay Kivel for useful discussions, Jan Bernauer for providing us with an updated version of the empirical TPE fit, Oleksii Gryniuk for providing us with the parametrizations of the photoabsorption cross section, and Carl Carlson for providing us with the parametrizations of the unpolarized proton structure functions. This work was supported by the Deutsche Forschungsgemeinschaft DFG in part through the Collaborative Research Center [The Low-Energy Frontier of the Standard Model (SFB 1044)], in part through the Graduate School [Symmetry Breaking in Fundamental Interactions (DFG/GRK 1581)], and in part through the Cluster of Excellence [Precision Physics, Fundamental Interactions and Structure of Matter (PRISMA)].


\begin{thebibliography}{99}  

\bibitem{Rosenbluth:1950yq} 
  M.~N.~Rosenbluth,
  Phys.\ Rev.\  {\bf 79}, 615 (1950).


\bibitem{Jones:1999rz} 
  M.~K.~Jones {\it et al.} [Jefferson Lab Hall A Collaboration],
  Phys.\ Rev.\ Lett.\  {\bf 84}, 1398 (2000)
  [nucl-ex/9910005].


\bibitem{Gayou:2001qd} 
  O.~Gayou {\it et al.} [Jefferson Lab Hall A Collaboration],
  Phys.\ Rev.\ Lett.\  {\bf 88}, 092301 (2002)
  [nucl-ex/0111010].


\bibitem{Punjabi:2005wq} 
  V.~Punjabi {\it et al.},
  Phys.\ Rev.\ C {\bf 71}, 055202 (2005)
  [Phys.\ Rev.\ C {\bf 71}, 069902 (2005)]
  [nucl-ex/0501018].


\bibitem{Puckett:2010ac} 
  A.~J.~R.~Puckett {\it et al.},
  Phys.\ Rev.\ Lett.\  {\bf 104}, 242301 (2010)
  [arXiv:1005.3419 [nucl-ex]].


\bibitem{Jones:2006kf} 
  M.~K.~Jones {\it et al.} [Resonance Spin Structure Collaboration],
  Phys.\ Rev.\ C {\bf 74}, 035201 (2006)
  [nucl-ex/0606015].


\bibitem{Guichon:2003qm} 
  P.~A.~M.~Guichon and M.~Vanderhaeghen,
  Phys.\ Rev.\ Lett.\  {\bf 91}, 142303 (2003)
  [hep-ph/0306007].


\bibitem{Blunden:2003sp} 
  P.~G.~Blunden, W.~Melnitchouk and J.~A.~Tjon,
  Phys.\ Rev.\ Lett.\  {\bf 91}, 142304 (2003)
  [nucl-th/0306076].


\bibitem{Carlson:2007sp} 
  C.~E.~Carlson and M.~Vanderhaeghen,
  Ann.\ Rev.\ Nucl.\ Part.\ Sci.\  {\bf 57}, 171 (2007)
  [hep-ph/0701272 [HEP-PH]].


\bibitem{Arrington:2011dn} 
  J.~Arrington, P.~G.~Blunden and W.~Melnitchouk,
  Prog.\ Part.\ Nucl.\ Phys.\  {\bf 66}, 782 (2011)
  [arXiv:1105.0951 [nucl-th]].


\bibitem{Chen:2004tw} 
  Y.~C.~Chen, A.~Afanasev, S.~J.~Brodsky, C.~E.~Carlson and M.~Vanderhaeghen,
  Phys.\ Rev.\ Lett.\  {\bf 93}, 122301 (2004)
  [hep-ph/0403058].


\bibitem{Borisyuk:2008es} 
  D.~Borisyuk and A.~Kobushkin,
  Phys.\ Rev.\ C {\bf 78}, 025208 (2008)
  [arXiv:0804.4128 [nucl-th]].


\bibitem{Borisyuk:2008db} 
  D.~Borisyuk and A.~Kobushkin,
  Phys.\ Rev.\ D {\bf 79}, 034001 (2009)
  [arXiv:0811.0266 [hep-ph]].


\bibitem{Kivel:2009eg} 
  N.~Kivel and M.~Vanderhaeghen,
  Phys.\ Rev.\ Lett.\  {\bf 103}, 092004 (2009)
  [arXiv:0905.0282 [hep-ph]].


\bibitem{Kivel:2012vs} 
  N.~Kivel and M.~Vanderhaeghen,
  JHEP {\bf 1304}, 029 (2013)
  [arXiv:1212.0683 [hep-ph]].


\bibitem{Meziane:2010xc} 
  M.~Meziane {\it et al.} [GEp2gamma Collaboration],
  Phys.\ Rev.\ Lett.\  {\bf 106}, 132501 (2011)
  [arXiv:1012.0339 [nucl-ex]].


\bibitem{Adikaram:2014ykv} 
  D.~Adikaram {\it et al.} [CLAS Collaboration],
  Phys.\ Rev.\ Lett.\  {\bf 114}, 062003 (2015)
  [arXiv:1411.6908 [nucl-ex]].


\bibitem{Rachek:2014fam} 
  I.~A.~Rachek {\it et al.},
  Phys.\ Rev.\ Lett.\  {\bf 114}, no. 6, 062005 (2015)
  [arXiv:1411.7372 [nucl-ex]].


\bibitem{Milner:2013daa} 
  R.~Milner {\it et al.} [OLYMPUS Collaboration],
  Nucl.\ Instrum.\ Meth.\ A {\bf 741}, 1 (2014)
  [arXiv:1312.1730 [physics.ins-det]].


\bibitem{Blunden:2005jv} 
  P.~G.~Blunden and I.~Sick,
  Phys.\ Rev.\ C {\bf 72}, 057601 (2005)
  [nucl-th/0508037].


\bibitem{Bernauer:2013tpr} 
  J.~C.~Bernauer {\it et al.} [A1 Collaboration],
  Phys.\ Rev.\ C {\bf 90}, no. 1, 015206 (2014)
  [arXiv:1307.6227 [nucl-ex]].


\bibitem{Borisyuk:2012he} 
  D.~Borisyuk and A.~Kobushkin,
  Phys.\ Rev.\ C {\bf 86}, 055204 (2012)
  [arXiv:1206.0155 [hep-ph]].


\bibitem{Borisyuk:2013hja} 
  D.~Borisyuk and A.~Kobushkin,
  Phys.\ Rev.\ C {\bf 89}, no. 2, 025204 (2014)
  [arXiv:1306.4951 [hep-ph]].


\bibitem{Lorenz:2014yda} 
  I.~T.~Lorenz, U.~G.~Mei§ner, H.-W.~Hammer and Y.-B.~Dong,
  Phys.\ Rev.\ D {\bf 91}, no. 1, 014023 (2015)
  [arXiv:1411.1704 [hep-ph]].


\bibitem{Borisyuk:2015xma} 
  D.~Borisyuk and A.~Kobushkin,
  Phys.\ Rev.\ C {\bf 92}, no. 3, 035204 (2015)
  [arXiv:1506.02682 [hep-ph]].


\bibitem{Tomalak:2014sva} 
  O.~Tomalak and M.~Vanderhaeghen,
  Eur.\ Phys.\ J.\ A {\bf 51}, no. 2, 24 (2015)
  [arXiv:1408.5330 [hep-ph]].


\bibitem{Zhou:2014xka} 
  H.~Q.~Zhou and S.~N.~Yang,
  Eur.\ Phys.\ J.\ A {\bf 51}, no. 8, 105 (2015)
  [arXiv:1407.2711 [nucl-th]].


\bibitem{Pohl:2010zza} 
  R.~Pohl {\it et al.},
  Nature {\bf 466}, 213 (2010).


\bibitem{Antognini:1900ns} 
  A.~Antognini {\it et al.},
  Science {\bf 339}, 417 (2013).


\bibitem{Mohr:2012tt} 
  P.~J.~Mohr, B.~N.~Taylor and D.~B.~Newell,
  Rev.\ Mod.\ Phys.\  {\bf 84}, 1527 (2012)
  [arXiv:1203.5425 [physics.atom-ph]].


\bibitem{Lee:2015jqa} 
  G.~Lee, J.~R.~Arrington and R.~J.~Hill,
  Phys.\ Rev.\ D {\bf 92}, no. 1, 013013 (2015)
  [arXiv:1505.01489 [hep-ph]].


\bibitem{Arrington:2015ria} 
  J.~Arrington and I.~Sick,
  J.\ Phys.\ Chem.\ Ref.\ Data {\bf 44}, 031204 (2015)
  [arXiv:1505.02680 [nucl-ex]].


\bibitem{Carlson:2015jba} 
  C.~E.~Carlson,
  Prog.\ Part.\ Nucl.\ Phys.\  {\bf 82}, 59 (2015)
  [arXiv:1502.05314 [hep-ph]].


\bibitem{Nevado:2007dd} 
  D.~Nevado and A.~Pineda,
  Phys.\ Rev.\ C {\bf 77}, 035202 (2008)
  [arXiv:0712.1294 [hep-ph]].


\bibitem{Hill:2011wy} 
  R.~J.~Hill and G.~Paz,
  Phys.\ Rev.\ Lett.\  {\bf 107}, 160402 (2011)
  [arXiv:1103.4617 [hep-ph]].


\bibitem{Carlson:2011zd} 
  C.~E.~Carlson and M.~Vanderhaeghen,
  Phys.\ Rev.\ A {\bf 84}, 020102 (2011)
  [arXiv:1101.5965 [hep-ph]].


\bibitem{Birse:2012eb} 
  M.~C.~Birse and J.~A.~McGovern,
  Eur.\ Phys.\ J.\ A {\bf 48}, 120 (2012)
  [arXiv:1206.3030 [hep-ph]].


\bibitem{Alarcon:2013cba} 
  J.~M.~Alarcon, V.~Lensky and V.~Pascalutsa,
  Eur.\ Phys.\ J.\ C {\bf 74}, no. 4, 2852 (2014)
  [arXiv:1312.1219 [hep-ph]].


\bibitem{Peset:2014jxa} 
  C.~Peset and A.~Pineda,
  Nucl.\ Phys.\ B {\bf 887}, 69 (2014)
  [arXiv:1406.4524 [hep-ph]].


\bibitem{McKinley:1948zz} 
  W.~A.~McKinley and H.~Feshbach,
  Phys.\ Rev.\  {\bf 74}, 1759 (1948).


\bibitem{Brown:1970te} 
  R.~W.~Brown,
  Phys.\ Rev.\ D {\bf 1}, 1432 (1970).


\bibitem{Gorchtein:2014hla} 
  M.~Gorchtein,
  Phys.\ Rev.\ C {\bf 90}, no. 5, 052201 (2014)
  [arXiv:1406.1612 [nucl-th]].


\bibitem{Drechsel:1997xv} 
  D.~Drechsel, G.~Knochlein, A.~Y.~Korchin, A.~Metz and S.~Scherer,
  Phys.\ Rev.\ C {\bf 57}, 941 (1998)
  [nucl-th/9704064].


\bibitem{Drechsel:1998zm} 
  D.~Drechsel, G.~Knochlein, A.~Y.~Korchin, A.~Metz and S.~Scherer,
  Phys.\ Rev.\ C {\bf 58}, 1751 (1998)
  [nucl-th/9804078].


\bibitem{Tarrach:1975tu} 
  R.~Tarrach,
  Nuovo Cim.\ A {\bf 28}, 409 (1975).


\bibitem{Gorchtein:2004ac} 
  M.~Gorchtein, P.~A.~M.~Guichon and M.~Vanderhaeghen,
  Nucl.\ Phys.\ A {\bf 741}, 234 (2004)
  [hep-ph/0404206].


\bibitem{Drechsel:2002ar} 
  D.~Drechsel, B.~Pasquini and M.~Vanderhaeghen,
  Phys.\ Rept.\  {\bf 378}, 99 (2003)
  [hep-ph/0212124].


\bibitem{Christy:2007ve} 
  M.~E.~Christy and P.~E.~Bosted,
  Phys.\ Rev.\ C {\bf 81}, 055213 (2010)
  [arXiv:0712.3731 [hep-ph]].


\bibitem{Arndt:2002xv} 
  R.~A.~Arndt, W.~J.~Briscoe, I.~I.~Strakovsky and R.~L.~Workman,
  Phys.\ Rev.\ C {\bf 66}, 055213 (2002)
  [nucl-th/0205067].

  
\end{thebibliography}
\end{document}